\documentclass[preprint,12pt]{elsarticle}

\usepackage{amssymb}
\usepackage{amsthm}
\usepackage{amsmath}

\usepackage{bm}
\usepackage{graphicx}
\usepackage{graphics}

\usepackage{subfigure}
\usepackage{subcaption}
\usepackage{tabularx,caption}
\usepackage{float} 
\usepackage{color}
\usepackage{multirow}
\usepackage{booktabs}
\usepackage{diagbox}
\usepackage{algorithm,algpseudocode}

\usepackage{lineno}
\usepackage{graphicx}

\usepackage{makecell}
\journal{XXX}

\begin{document}

\begin{frontmatter}



\title{Scalable physics-informed deep generative model for solving forward and inverse stochastic differential equations}

\author[hust]{Shaoqian Zhou}
\author[hust]{Wen You}
\author[shnu]{Ling Guo}
\author[hust]{Xuhui Meng \fnref{1}}

% \affiliation[inst2]{organization={Department Two},%Department and Organization
%             addressline={Address Two}, 
%             city={City Two},
%             postcode={22222}, 
%             state={State Two},
%             country={Country Two}}

\address[hust]{Institute of Interdisciplinary Research for Mathematics and Applied Science, School of Mathematics and Statistics, Huazhong University of Science and Technology, Wuhan 430074, China}
\address[shnu]{Department of Mathematics, Shanghai Normal University, Shanghai, China}

\fntext[1]{Corresponding author: xuhui\_meng@hust.edu.cn (Xuhui Meng).}
% \author[inst1]{Author One}

% \affiliation[inst1]{organization={Department One},%Department and Organization
%             addressline={Address One}, 
%             city={City One},
%             postcode={00000}, 
%             state={State One},
%             country={Country One}}

% \author[inst2]{Author Two}
% \author[inst1,inst2]{Author Three}

% \affiliation[inst2]{organization={Department Two},%Department and Organization
%             addressline={Address Two}, 
%             city={City Two},
%             postcode={22222}, 
%             state={State Two},
%             country={Country Two}}

\begin{abstract}
Physics-informed deep learning approaches have been developed to solve forward and inverse stochastic differential equation (SDE) problems with high-dimensional stochastic space. However, the existing deep learning models have difficulties solving SDEs with high-dimensional spatial space. 
In the present study, we propose a scalable physics-informed deep generative model (sPI-GeM), which is capable of solving SDE problems with both high-dimensional stochastic and spatial space. The sPI-GeM consists of two deep learning models, i.e., (1) physics-informed basis networks (PI-BasisNet), which are used to learn the basis functions as well as the coefficients given data on a certain stochastic process or random field, and (2) physics-informed deep generative model (PI-GeM), which learns the distribution over the coefficients obtained from the PI-BasisNet. The new samples for the learned stochastic process can then be obtained using the inner product between the output of the generator and the basis functions from the trained PI-BasisNet. The sPI-GeM addresses the scalability in the spatial space in a similar way as in the widely used dimensionality reduction technique, i.e., principal component analysis (PCA). A series of numerical experiments, including approximation of Gaussian and non-Gaussian stochastic processes, forward and inverse SDE problems, are performed to demonstrate the accuracy of the proposed model. Furthermore, we also show the scalability of the sPI-GeM using an example of a forward SDE problem with both high-dimensional stochastic and spatial space, respectively.   
\end{abstract}


% \begin{highlights}
% \end{highlights}


\end{frontmatter}


\section{Introduction}
\label{sec:introduction}
Stochastic differential equations (SDEs) are differential equations involving uncertain coefficients and/or random forcing terms/boundary/initial conditions, which result in uncertainties in the sought solutions \cite{bjork2004martingale,braumann2019introduction,psaros2023uncertainty,patil2020real,patil2023reduced,yang2012adaptive,wang2024accelerating}.  To quantify uncertainties in SDEs, various numerical methods have been developed. Specifically, the Monte Carlo (MC) method and the generalized polynomial chaos (gPC) are two of the most popular approaches among the existing numerical solvers. As pointed out in \cite{zhong2023pi}, the former is robust and straightforward, but in general comes with expensive computational cost. In addition,  the latter is computationally more efficient but suffers from the ``curse of dimensionality'' (CoD).

Recently, deep learning has achieved remarkable progress in solving both forward and inverse partial differential equations (PDEs) \cite{yu2018deep,sirignano2018dgm,raissi2019physics,fu2024physics}, especially for high-dimensional PDEs \cite{han2018solving,zang2020weak,hu2024score,hu2024tackling,mao2023physics,guo2022monte} since the deep neural networks (DNNs) are capable of breaking the ``curse of dimensionality''. In particular, the physics-informed neural networks (PINNs) \cite{raissi2019physics,mao2023physics,zou2025learning,xiang2025physics} are one of the most widely used deep learning methods for solving PDEs due to their effectiveness as well as straightforward implementations.   Specifically,  PINNs employ DNNs to approximate the solution to a given PDE, and then the automatic differentiation is utilized to encode the corresponding PDE to DNNs. For deterministic PDEs, we can train the PINNs by minimizing the mean squared errors (MSE) of the residual for the equations and the mismatches between the PINN predictions and the observational data.  In addition to deterministic PDEs, variants of PINNs have also been proposed to solve SDEs \cite{psaros2023uncertainty,zou2024neuraluq}.  For instance, Zhang {\sl et al} proposed to combine the PINNs with the arbitrary polynomial chaos (NN-aPC) to solve both the forward and inverse SDE problems \cite{zhang2019quantifying}. Although effective, it is still challenging for the NN-aPC to handle high-dimensional problems because the number of polynomial chaos terms grows exponentially as the dimension increases \cite{guo2022normalizing}. Inspired by the capability of deep generative models for handling high-dimensional data, numerous physics-informed deep generative models (PI-GeMs) have been developed to solve high-dimensional SDEs, e.g., physics-informed generative adversarial networks (PI-GANs) \cite{yang2020physics,yang2019adversarial}, physics-informed variational autoencoder (PI-VAE) \cite{zhong2023pi,shin2023physics}, and physics-informed normalizing flows \cite{guo2022normalizing}, to name just a few. Numerical results on SDEs with 100 stochastic dimensions have been reported in \cite{yang2020physics}, which are quite challenging for the conventional numerical methods to handle. 

% Although significant progress has been made on solving SDEs with high dimensions in stochastic domain using the PI-GeMs, there are still limitations to be addressed in the existing models. For the approach in \cite{shin2023physics}, the posterior distribution in the autoencoder is factorized as the diagonal Gaussian distribution, which limits the expressivity of the models.

Although significant progress has been made on solving SDEs with high dimensions in the stochastic domain using the PI-GeMs, most of the existing models have difficulties scaling to SDE problems with high-dimensional physical space, e.g., spatial, or temporal-spatial space, and so on. To the best of our knowledge, results on SDE problems with spatial dimensions greater than two have not been reported in the existing works using deep learning \cite{shin2023physics,zhong2023pi,yang2020physics}. In general, the deep generative models are designed to approximate unknown distributions given empirical samples.  To approximate stochastic processes using deep generative models, a commonly employed approach is to represent each sample from the target stochastic process using numbers of discrete points. The target stochastic process can then be treated as an unknown distribution with the dimensionality equal to the number of points used to resolve each sample. For the training of deep generative models,  we can minimize a certain metric that is able to measure the dissimilarity between the generated and the target stochastic process, e.g., the maximum mean discrepancy (MMD) in \cite{zhong2023pi}, the Kullback–Leibler (KL) divergence \cite{guo2022normalizing}, and the Wasserstein-1 ($\bm{W}$-1) distance in \cite{yang2020physics}. In addition, the metric is estimated empirically based on the generated and observed samples. For problems with spatial dimensions greater than two, thousands of points are required to accurately resolve each sample, leading to prohibited computational cost in the estimation of the metric. In other words, it is challenging for the aforementioned models to handle SDEs with high-dimensional spatial domains.

{
Note  that stochastic processes or SDEs defined on high-dimensional physical space arise naturally in several real-world applications. For instance, the Schr$\ddot{\mbox{o}}$dinger equation, in which the physical dimensionality scales as $3N$ with $N$ denoting the number of particles, is widely used to model particle dynamics in ordered solids. In disordered solids,  however, the random Schr$\ddot{\mbox{o}}$dinger operator, which incorporates a random potential, is generally employed for more accurate descriptions of particle dynamics \cite{anderson1958absence,kirsch2007invitation,denisov2005spectral}. It leads to SDEs posed on high-dimensional physical domains. More recently, there has also been interest in inferring the thermal conductivity of solids by solving the inverse phonon Boltzmann transport equation with uncertainties in micro-/nano-scale heat conduction \cite{li2025physics}. This approach requires assigning a stochastic process with 7 physical dimensions (i.e., 1 temporal, 3 angular and 3 spatial dimensions) as the prior for the energy density function. Hence, numerical solvers for solving SDEs that are scalable in both stochastic and physical dimensions are still desirable, since the existing methods struggle with problems involving high-dimensional spatial domains, as previously mentioned.}

The primary contributions of this work are listed as follows:
(1) We develop a scalable physics-informed deep generative model (sPI-GeM),  which is capable of efficiently solving both forward and inverse SDE problems, and is also able to handle high-dimensional problems in both the stochastic and spatial domains. (2) We perform numerical experiments on solving forward/inverse SDE problems with high dimensions in both stochastic ($> 50$ dimensions) and spatial (20 dimensions) space. In particular, the SDE problem with 20 dimensional spatial space considered in the present study has not been reported in the existing work to the best of our knowledge.

The rest of this paper is organized as follows: In Sec. \ref{sec:method}, we present the problem formulation as well as the scalable physics-informed deep generative models for solving stochastic differential equations; the numerical results are shown in Sec. \ref{sec:results}, and a summary on this study is present in Sec. \ref{sec:summary}.

\section{Methodology}
\label{sec:method}

\subsection{Problem formulation}

Consider a general steady stochastic differential equation (SDE) for the dynamics of a physical system as follows:

\begin{subequations}\label{eq:spde}
\begin{equation}
\mathcal{N}_{\lambda, \bm{\zeta}}[u(\bm{x}, \bm{\zeta})] = f(\bm{x}, \bm{\zeta}), ~\bm{x} \in \Omega, ~\bm{\zeta} \in Z, \\
\label{eq:spde_a}
\end{equation}
\begin{equation}
\mathcal{B}_{\lambda, \bm{\zeta}} [u(\bm{x}_{bc}, \bm{\zeta})] = b(\bm{x}_{bc}, \bm{\zeta}), ~ \bm{x}_{bc} \in \Gamma,
\label{eq:spde_b}
\end{equation}
\end{subequations}
where $\bm{x}$ represents the $D_{\bm{x}}$-dimensional space coordinate,  $\bm{x}_{bc}$ denotes the coordinate at the boundary, $\bm{\zeta}$ is a $D_{\bm{\zeta}}$-dimensional random variable in a probability space $Z$,  $u$ is the solution to Eq. \eqref{eq:spde}, $\mathcal{N}$ denotes any operator, e.g., linear or nonlinear differential operator parametrized by $\lambda$,  $f$ is the forcing term, which is either random or deterministic related to the specific problem at hand,  $\mathcal{B}$ is the operator imposed on the boundaries, $b$ is the boundary condition, and $\Omega$ is a bounded domain with the boundary $\Gamma$.

Similar as in \cite{yang2020physics,shin2023physics}, we consider two particular problems: (1) {\emph{forward problem}}: $\lambda$ or the operator $\mathcal{N}$ is known, $f$ and/or $b$ are represented by given data,  and we would like to seek the solution to Eq. \eqref{eq:spde}; and (2) \emph{inverse or mixed problem}:  $\lambda$ is an unknown parameter or field, and we have partial observations on $u/\lambda/f/b$. The objective is then to obtain the predictions on the solution $u$ as well as the unknown $\lambda$.

% \subsection{Scalable deep genrative model for stochastic process}

\subsection{Scalable physics-informed deep generative model}

In this subsection, we first introduce the overview of the proposed scalable physics-informed deep generative model (sPI-GeM), and then present the details on the training of this model for solving the SDE problems.

\begin{figure}[h]
    \centering
    \subfigure[Physics-informed basis networks (PI-BasisNet) ]{\label{fig:spi-gem-a}
    \includegraphics[width=0.9\textwidth]{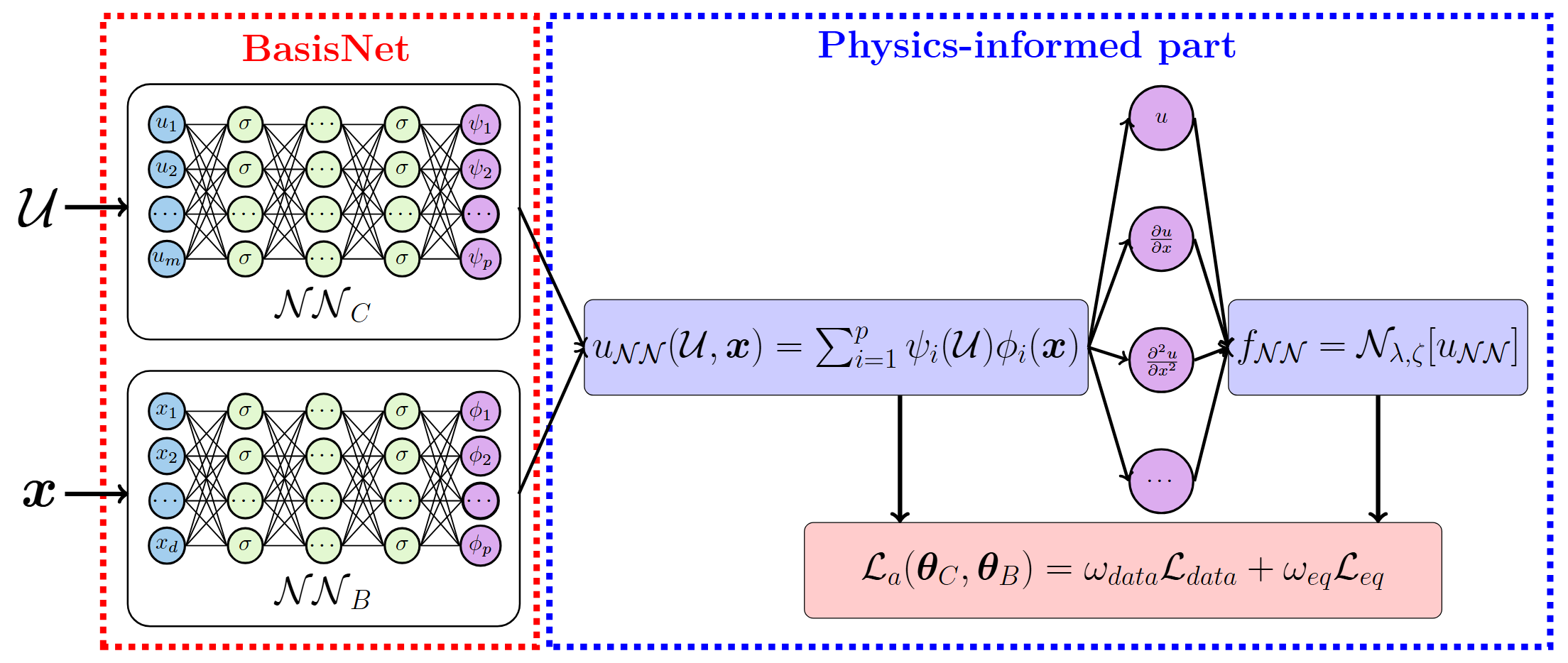}
    }
    \subfigure[Physics-informed generative model (PI-GeM) ]{\label{fig:spi-gem-b}
    \includegraphics[width=0.9\textwidth]{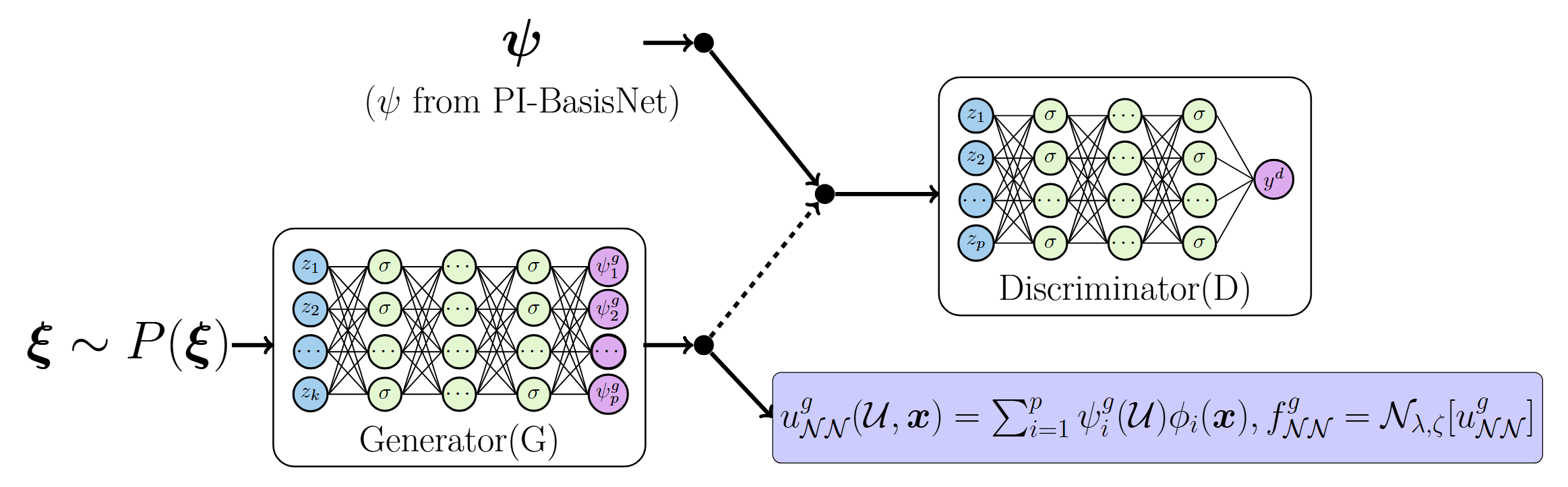}}
    \caption{\label{fig:spi-gem}
    Schematic of scalable physics-informed deep generative models (sPI-GeM) for solving SDE problems, which consists of two different deep learning models: (a) Physics-informed basis networks (PI-BasisNet), and (b) Physics-informed deep generative model (PI-GeM).}
\end{figure}

As shown in Fig. \ref{fig:spi-gem}, the sPI-GeM consists of two different deep learning models, i.e. physics-informed basis networks (PI-BasisNet, Fig. \ref{fig:spi-gem-a}) and physics-informed deep generative model (PI-GeM, Fig. \ref{fig:spi-gem-b}). In the PI-BasisNet, there are two subnetworks, i.e., $\mathcal{NN}_C(\mathcal{U}; \bm{\theta}_C)$ and $\mathcal{NN}_B(\bm{x}; \bm{\theta}_B)$. The first subnetwork $\mathcal{NN}_C$ is parameterized by $\bm{\theta}_{C}$. It takes as input $\mathcal{U}$ and outputs $\bm{\psi}$ where $\bm{\psi} = (\psi_1, ..., \psi_p)^T$ with $p$ denoting the number of dimensions for $\bm{\psi}$. Note that $\mathcal{U}$ is different for different problems, which will be clarified in the last part of this subsection as well as each case in Sec. \ref{sec:results}. In addition,  $\mathcal{NN}_B$ is parameterized by $\bm{\theta}_{B}$ and takes as input the spatial coordinate $\bm{x}$. The output of $\mathcal{NN}_B$ is $\bm{\phi}$ where $\bm{\phi} = (\phi_1, ..., \phi_p)^T$. As for the BasisNet, its output is obtained using the inner product of $\bm{\psi}(\mathcal{U})$ and $\bm{\phi}(\bm{x})$ as follows:
\begin{equation}\label{eq:basis_output}
u_{\mathcal{NN}}(\mathcal{U}, \bm{x}) = \sum^p_{i=1} \psi_i(\mathcal{U}) \phi_i(\bm{x}).
\end{equation}
Also, we can encode the differential equations to the BasisNet using the automatic differentiation \cite{raissi2019physics}, which yields the PI-BasisNet as illustrated in Fig. \ref{fig:spi-gem-a}.  Further, the generative model used in this study is the generative adversarial networks (GANs) \cite{goodfellow2014generative,arjovsky2017wasserstein,gulrajani2017improved}, which have a generator ($G(\bm{\xi}; \bm{\theta}_G) = \mathcal{NN}_G(\bm{\xi}; \bm{\theta}_G)$)  and a discriminator ($D(\cdot; \bm{\theta}_G) = \mathcal{NN}_D(\cdot, \bm{\theta}_D)$), where $\bm{\theta}_G$ and $\bm{\theta}_D$ denote the trainable parameters in the generative model. Specifically, the generator $\mathcal{NN}_G$ takes as input the samples drawn from the standard normal distribution, i.e. $P(\bm{\xi})$, and the output of $\mathcal{NN}_G$ is the approximation to $\bm{\psi}(\mathcal{U})$ that are obtained from the trained PI-BasisNet. In other words, the generator is utilized to learn the distribution over $\bm{\psi}$ using the samples from the trained PI-BasisNet. The discriminator $\mathcal{NN}_{D}$ takes the target data and the generated samples from $\mathcal{NN}_G$ as input and outputs the metric to measure the dissimilarity between them.  Here we note that the BasisNet and PI-BasisNet share the same architectures as the functional prior \cite{meng2022learning,zou2023hydra}/deep operator networks (DeepONet) \cite{lu2021learning} and the physics-informed deep operator network (PI-DeepONet)  \cite{wang2021learning}, respectively.  We refer to this architecture as PI-BasisNet in the present study to avoid any confusion since we do not aim to learn the functional prior or operator between functions.

We now discuss the training of sPI-GeM for solving SDE problems. For forward problems, we assume that we have $N_{\mathcal{F}} ~(\mathcal{F} = f,~ b)$ snapshots for $f$ as well as the boundary conditions $b$. Each snapshot of $f$ or $b$ is represented by numbers of measurements from the corresponding sensors, which are expressed as follows:
\begin{equation}\label{eq:sde_data}
\mathcal{D}_\mathcal{F} = \{\mathcal{F}^j(\bm{x}^j_1),  ..., \mathcal{F}^j(\bm{x}^j_{N^j_{\mathcal{F}}})\}^{N_{\mathcal{F}}}_{j=1}, ~\mathcal{F} = f,~b,
\end{equation}
where $\bm{x}^j_i~ (i = 1, ...,  N^j_{\mathcal{F}})$ and $N^j_{\mathcal{F}}$ denote the locations of the measurements and the number of sensors for the $jth$ snapshot i.e. $\mathcal{F}^j$. The output of BasisNet i.e. $u_{\mathcal{NN}}$ is to approximate the solution to Eq. \eqref{eq:spde},  and we can minimize the following loss function to train the PI-BasisNet: 
\begin{equation}\label{eq:loss_step1}
\mathcal{L}_a(\bm{\theta}_C, \bm{\theta}_B) = \omega_{data}\mathcal{L}_{data} + \omega_{eq} \mathcal{L}_{eq},
\end{equation}
where 
\begin{equation}\label{eq:loss_basis}
\begin{aligned}
\mathcal{L}_{data} &= \frac{1}{N_{b}}\sum^{N_b}_{j=1}\left[\frac{1}{N^j_{bc}}\sum^{N^j_{bc}}_{i=1}|\mathcal{B}_{\lambda, \zeta^j}[u^j_{\mathcal{NN}}(\mathcal{U}^j, \bm{x}^j_{bc,i})] - b^j(\bm{x}^j_{bc,i})|^2 \right], \\
\mathcal{L}_{eq} &=  \frac{1}{N_f}\sum^{N_f}_{j=1}\left[\frac{1}{N^j_{eq}}\sum^{N^j_{eq}}_{i=1}|R^j_i|^2\right], ~ R^j_i = \mathcal{N}_{\lambda, \zeta_i} [u_{\mathcal{NN}}(\mathcal{U}^j, \bm{x}^j_i)] - f^j(\bm{x}^j_i).
\end{aligned}
\end{equation}
In Eq. \eqref{eq:loss_step1}, $\omega_{data}$ and $\omega_{eq}$ are the weights to balance each loss term in the loss function.  For the $jth$ snapshot of $f$ and $b$ in Eq. \eqref{eq:loss_basis}: (1) $N^j_{bc}$ and $N^j_{eq}$ are the numbers of points for evaluating the losses for boundary condition and the residual of Eq. \eqref{eq:spde}, respectively; (2) $\mathcal{U}^j$ is the representation for $f$ and/or $b$, which can be obtained by using the representation of $f$ or concatenating the representations of $f$ and $b$ in this study; and (3) $R^j_i$  is the residual of Eq. \eqref{eq:spde} at the location $\bm{x}^j_i$.  Upon the training of PI-BasisNet, we can obtain the predictions for $u$ corresponding to $\mathcal{D}_{\mathcal{F}}$ at any location in the computational domain. {In general, we would like to seek a stochastic process for $u$ given $f/b$ in solving SDEs. With the trained PI-BasisNet, however, we can only obtain samples of $u$ related to each $f/b$ in the training dataset instead of a stochastic process. We then propose to obtain the stochastic process for $u$ using the deep generative model in Fig. \ref{fig:spi-gem-b}.}

{
Note that in the conventional methods for solving stochastic problems, e.g., the polynomial chaos expansion (PCE) and Karhunen–Lo$\grave{\mbox{e}}$ve expansion (KLE), the solution to a SDE is in general expressed as the inner product of random coefficients and the deterministic basis functions.  In the current study, we also express the solution to Eq. \eqref{eq:spde} in a similar way, i.e.,
\begin{equation}\label{eq:process_u}
u(\bm{x}, \bm{\zeta}) = \sum^{p_u}_{i=1} \psi_{u,i}(\bm{\zeta}) \phi_{u,i}(\bm{x}), 
\end{equation}
where $\psi_{u,i}(\bm{\zeta})$ and $\phi_{u,i}(\bm{x})$ denote the random coefficients and the basis functions, respectively.  With the trained PI-BasisNet at the first stage, we are able to obtain a set of basis functions $\bm{\phi}(\bm{x})$ which are expressive enough to represent $u$ corresponding to each $f/b$ in the training dataset. We propose to approximate the solution to  Eq. \eqref{eq:spde} as
\begin{equation}
u(\bm{x}, \bm{\zeta}) \approx \sum^{p}_{i=1} \psi_{i}(\bm{\zeta}) \phi_{i}(\bm{x}),
\end{equation}
where (1) $\bm{\phi}(\bm{x})$ is from the trained PI-BasisNet, and (2) $\psi_{i}(\bm{\zeta})$ are samples from a certain hidden distribution $P_{\psi,i}$. Up to now, the stochastic process $u(\bm{x}, \bm{\zeta})$ can be determined by ${\psi}_{i}(\bm{\zeta})$ as we assume that $\bm{\phi}(\bm{x})$ are linearly independent. Further, the coefficients $\bm{\psi}(\mathcal{U})$ for each $u$ corresponding to $f/b$ in the training dataset can also be obtained in the trained PI-BasisNet, which are viewed as realizations  of ${\psi}_{i}(\bm{\zeta})$ or samples from $P_{\psi,i}$. Approximating $u(\bm{x}, \bm{\zeta})$ is now switched to learn the hidden distributions $P_{\psi,i}$ given samples for ${\psi}_{i}(\bm{\zeta})$, which is then achieved using deep generative models. As aforementioned in Sec. \ref{sec:introduction}, most of the existing deep generative models have difficulties handling SDE problems with high-dimensional spatial space, since a large amount of discrete points are required to accurately resolve each sample in the computation of the employed metric or loss function.  In the present method, however, we utilize the deep generative models to learn the distributions over the coefficients ${\psi}_{i}(\bm{\zeta})$ for $u$, which has a dimension of $p$, i.e., the number of basis functions. Generally, the number of the basis functions is much less than the number of discrete points to resolve $u$, especially in high-dimensional problems. We can therefore address the  scalability issue in the spatial space of the existing deep generative models for solving stochastic problems.
Furthermore, we can also expect faster convergence for the present model due to the dimensionality reduction compared to the existing  models, which will be shown in Sec. \ref{sec:part_1}.
% Furthermore, as we discussed in Sec. \ref{sec:introduction}, most of the existing deep generative models have difficulties handling SDE problems with high-dimensional spatial space, since a large amount of discrete points are required to accurately resolve each sample in the computation of the employed metric or loss function.  In the present method, we utilize the coefficients $\bm{\psi}(\mathcal{U})$ to represent $u$. Generally, the number of the basis functions is much less than the number of discrete points to resolve $u$, especially in high-dimensional problems. We can therefore address the  scalability issue in the spatial space of the existing deep generative models.
% Furthermore, we can also expect faster convergence for the present model due to the dimensionality reduction compared to the existing generative models, which will be shown in Sec. \ref{sec:part_1}.
}

% we then have access to the coefficients of each sample $u$ in the training data.  {\color{red}  The coefficients $\bm{\psi}(\mathcal{U})$ are utilized as the compact representation for $u$, rather than the numbers of discrete points that are widely used in the existing models.  Once we have the  representations for each snapshot of $u$ in the training dataset, we can approximate the distribution over $\bm{\psi}(\mathcal{U})$ using the deep generative models.} Generally, the number of the basis functions is much less than the number of discrete points to resolve $u$, especially in high-dimensional problems. We can therefore address the issue of unscalability in the spatial space of the existing deep generative models.
% Furthermore, we can also expect faster convergence for the present model due to the dimensionality reduction compared to the existing generative models, which will be shown in Sec. \ref{sec:part_1}. 

% We would like to discuss that the present approach shares similarity with the polynomial chaos expansion (PCE) or Karhunen-Lo$\grave{\mbox{e}}$ve expansion (KLE) that are widely used in solving SDEs, i.e., the stochastic process is approximated by the inner product of the coefficients and basis functions. More specifically, the coefficients and basis functions in the present model are learned given training  data using the generative model and the PI-BasisNet, respectively.  Although it does not have the solid mathematical foundation as in the PCE/KLE yet, the present model is more flexible than the PCE and KLE due to the expressivity of the DNNs.

The particular generative model used in the present study is the Wasserstein generative adversarial networks with gradient penalty (WGAN-GP) \cite{gulrajani2017improved}, which is a universal approximator to any distribution \cite{lu2020universal} and is also efficient for generating samples. We minimize the following loss functions to train the generator and the discriminator of WGAN-GP in an alternative way \cite{gulrajani2017improved,yang2020physics,meng2022learning}:
\begin{equation} \label{eqn:GDloss}
\begin{aligned}
\mathcal{L}_G =& -\mathbb{E}_{\bm{\xi}\sim P(\bm{\xi})} [ D(G(\bm{\xi}; \bm{\theta}_G); \bm{\theta}_D)], \\
\mathcal{L}_D = &~ \mathbb{E}_{\bm{\xi}\sim P(\bm{\xi})} [ D(G(\bm{\xi}; \bm{\theta}_G); \bm{\theta}_D)] - \mathbb{E}_{\bm{\psi} \sim P_{\bm{\psi}}} [ D(\bm{\psi}; \bm{\theta}_D)] +\\
&  ~ \omega \mathbb{E}_{\hat{T} \sim P_i} (\Vert \nabla_{\hat{T}} D(\hat{T}; \bm{\theta}_D) \Vert_2 -1 )^2,
\end{aligned}
\end{equation}
where $\bm{\psi}$ is from the trained PI-BasisNet, $P_i$ is the distribution induced by uniform sampling on interpolation lines between independent samples of $\bm{\psi}$ and $G(\bm{\xi};  \bm{\theta}_G)$ \cite{gulrajani2017improved,yang2020physics,meng2022learning}, and $\omega$ is the gradient penalty coefficient. As mentioned in \cite{meng2022learning}, the loss function for the generator can be interpreted as the {$\bm{W}$} - 1 distance between the generated and the target distribution, up to constants. 
{More details on the training of WGAN-GP is illustrated in Algorithm \ref{alg:wgan-gp}. Note that all the expectations in Eq. \eqref{eqn:GDloss} are computed using the Monte Carlo method based on $B_{\bm{\xi}}$ samples at each training step.}

\begin{algorithm}[H]
\caption{Details for training the WGAN-GP in sGeM/sPI-GeM.}
\label{alg:wgan-gp}
\begin{algorithmic}
\Require 
\begin{quote}
\hspace{-1.cm} 
\begin{itemize}
    \item Samples for $\bm{\psi}$ from the trained BasisNet/PI-BasisNet as well as the batch size $B_{\bm{\xi}}$ used in minibatch training. \;
    \item The number of training steps $n_g$ for the generator, the gradient penalty coefficient $\omega$,  and the number of iterations for the discriminator $S_D$  per iteration of the generator. In particular, $\omega = 10$ and $S_D = 10$ in this study. \;
\end{itemize}
\end{quote}

\For{$k_g = 1, ..., n_g$}

\For{$k_d = 1, ..., S_D$}
\begin{quote}
1. Sample $\{\bm{\xi}^{(j)}\}_{j=1}^{B_{\bm{\xi}}}$ independently from $\mathcal{N}(\bm{0}, \bm{I}_{D_{\bm{\xi}}})$. \;

2. Sample $\{\bm{\psi}^{(j)}\}_{j=1}^{B_{\bm{\xi}}}$ independently from $\{\bm{\psi}\}_{j=1}^{N_{\mathcal{F}}},~ \mathcal{F} = u, \lambda$. \;

3. Sample $\{{\epsilon}^{(j)}\}_{j=1}^{B_{\bm{\xi}}}$ independently from a uniform distribution $\mathcal{U}[0, 1]$.\;

4. Compute $\{\hat{T}_j\}^{B_{\bm{\xi}}}_{j=1}$ based on $\hat{T}_i = \epsilon_j {\bm{\psi}_j} + (1 - \epsilon_j) G(\bm{\xi}_j; \bm{\theta}_G)$.\;

5. Update  $\bm{\theta}_D$ based on Eq. \eqref{eqn:GDloss} using the Adam optimizer.\;
\end{quote}
\EndFor

\begin{quote}
1. Sample $\{\bm{\xi}^{(j)}\}_{j=1}^{B_{\bm{\xi}}}$ independently from $\mathcal{N}(\bm{0}, \bm{I}_{D_{\bm{\xi}}})$. \;

% 2. Sample $\{\bm{\psi}^{(j)}\}_{j=1}^{B_{\bm{\psi}}}$ independently from $\{\bm{\psi}\}_{j=1}^{N_{\mathcal{F}}},~ \mathcal{F} = u, \lambda$. \;

2. Update $\bm{\theta}_G$ based on Eq. \eqref{eqn:GDloss} using the Adam optimizer.\;
\end{quote}

\EndFor
\\
\end{algorithmic}
\end{algorithm}

Upon the training of the deep generative model, we can then generate samples for $u$ and $f$ as follows:
\begin{equation}\label{eq:output_sgem}
\begin{aligned}
u_G(\bm{\xi}; \bm{x}) = \sum^p_{i=1}G_i(\bm{\xi}; \bm{\theta}_G) \phi_i(\bm{x}),
~f_G(\bm{\xi}; \bm{x}) = \mathcal{N}_{\lambda, \zeta}[u_G(\bm{\xi}; \bm{x})], 
\end{aligned}
\end{equation}
where $G(\bm{\xi}; \bm{\theta}_G)$ is the output of the generator with $\bm{\xi} \sim P(\bm{\xi})$ as the input, $\phi_i(\bm{x})$ is from the trained PI-BasisNet and $f_G$ is computed via the automatic differentiation as in the PI-BasisNet.  {
We would like to note that the proposed approach shares similarities with the conventional methods such as PCE and KLE, both of which are widely used in solving stochastic problems. In these methods, a stochastic process is expressed as an inner product  of deterministic basis functions and random coefficients. The key distinction lies in how these components are obtained: in PCE and KLE, the basis functions are fixed a priori, e.g., orthogonal polynomials or eigenfunctions of a covariance operator, while  the coefficients are determined from data or problem parameters. In contrast, our method learns both the basis functions and the coefficients from the training data using DNNs. Specifically, the coefficients are approximated by a generative model, and the basis functions are adaptively learned via the BasisNet/PI-BasisNet. We acknowledge that the current framework does not yet possess the same rigorous mathematical foundation as PCE or KLE, particularly regarding convergence guarantees, but it benefits from the expressive power of DNNs. This flexibility allows the model to represent complex, non-Gaussian, and high-dimensional stochastic processes that are challenging to handle with existing methods.}

% As aforementioned in Sec. \ref{sec:introduction}, most of the existing deep generative models have difficulties handling SDE problems with high-dimensional spatial space, since a large amount of discrete points are required to accurately resolve each sample in the computation of the employed metric or loss function.  In the present method, we utilize the deep generative models to learn the distribution over the compact representation of $u$, i.e., $\bm{\psi}(\mathcal{U})$ rather than using numbers of discrete points that is widely used in the existing models. Generally, the number of the basis functions is much less than the number of discrete points to resolve $u$, especially in high-dimensional problems. We can therefore address the  scalability issue in the spatial space of the existing deep generative models.
% Furthermore, we can also expect faster convergence for the present model due to the dimensionality reduction compared to the existing generative models, which will be shown in Sec. \ref{sec:part_1}.

For inverse or mixed problems, we have snapshots for $u$ and possibly $\lambda$ in addition to $f$ and/or $b$.  For the case where $\lambda$ is a random coefficient, we use partial outputs of the $\mathcal{NN}_C$ to approximate it in the PI-BasisNet. While for the case where $\lambda$ is a random field,  we utilize another neural network for the basis functions of $\lambda$, and divide the outputs of $\mathcal{NN}_C$ into two parts for the coefficients of $u$ and $\lambda$, respectively. As for the training of PI-BasisNet, we add the following losses for $u$ and possibly $\lambda$ in $\mathcal{L}_{data}$ of Eq. \eqref{eq:loss_basis}, and keep the loss for the equation unchanged. The additional loss is expressed as:
\begin{equation}
\mathcal{L}_{data,\mathcal{F}} = \frac{1}{N_{\mathcal{F}}}\sum^{N_{\mathcal{F}}}_{j=1}\left[\frac{1}{N^j_{\mathcal{F}}}\sum^{N^j_{\mathcal{F}}}_{i=1}|\mathcal{F}^j_{\mathcal{NN}}(\mathcal{U}^j, \bm{x}^j_{i}) - \mathcal{F}^j(\bm{x}^j_{i})|^2 \right], ~ \mathcal{F}  = u,~\lambda.
\end{equation}
Further, the loss function for training the PI-GeM is the same as in the forward problems. With the trained PI-GeM, we can obtain the samples for $\lambda$ in a similar way to obtain $u$, as shown in Eq. \eqref{eq:output_sgem}.  A summary on the sPI-GeM for solving SDE problems is illustrated in Algorithm \ref{alg:spi-gem}.

\begin{algorithm}[H]
\caption{sPI-GeM for solving SDE problems}
\label{alg:spi-gem}
\begin{algorithmic}
\Require 
\begin{quote}
\hspace{-1.cm} 
\begin{itemize}
    \item Training data on $f$ and $b$, i.e., $\mathcal{D}_f$ and $\mathcal{D}_b$ in forward problems, or training data on $f$, $u$ and possibly $\lambda$, i.e., $\mathcal{D}_f$, $\mathcal{D}_u$, and/or $\mathcal{D}_{\lambda}$, for inverse or mixed problems. 
    \item Batch sizes for training PI-BasisNet and PI-GeM, i.e., $B_{\mathcal{F}}$ and $B_{\bm{\xi}}$, respectively.
\end{itemize}
\end{quote}

\\
\item[{\bf{Step I}}]: PI-BasisNet for learning basis functions

\For{$k=1,2...T_1$}

\begin{quote}
1. Sample $\{\mathcal{F}^{j}\}_{j=1}^{B_{\mathcal{F}}}$ independently from $\mathcal{D}_{\mathcal{F}}$, $\mathcal{F} = f, b$.\;

2. Perform one gradient descent step to update $\bm{\theta}_C$, $\bm{\theta}_B$, and/or the parameters to parameterize $\lambda$ based on Eq. \eqref{eq:loss_step1} using Adam optimizer.\;
\end{quote}
\EndFor

\\
\item[{\bf{Step II}}]: PI-GeM for approximating stochastic process

\For{$k=1,2...T_2$}

\begin{quote}
1. Sample $\{\bm{\xi}^{(j)}\}_{j=1}^{B_{\bm{\xi}}}$ independently from $\mathcal{N}(\bm{0}, \bm{I}_{D_{\bm{\xi}}})$. \;

2. Sample $\{\bm{\psi}^{(j)}\}_{j=1}^{B_{\bm{\psi}}}$ independently from $\{\bm{\psi}\}_{j=1}^{N_{\mathcal{F}}},~ \mathcal{F} = u, \lambda$. \;

3. Update $\bm{\theta}_G$ and $\bm{\theta}_D$ alternately based on Eq. \eqref{eqn:GDloss} using the Adam optimizer.\;
\end{quote}
\EndFor

\\
\item[{\bf{Step III}}]: Predictions from sPI-GeM

\begin{quote}
Generate samples for $u$, $f$ and $\lambda$ from the trained sPI-GeM based on Eq. \eqref{eq:output_sgem}.\;
\end{quote}

\end{algorithmic}
\end{algorithm}

Finally, we discuss the treatment of the input for  $\mathcal{NN}_C$ in the PI-BasisNet, especially for high-dimension problems. Without loss of generality, we assume that $\mathcal{U}$ is the representation of $f$. A straightforward way is to use measurements from sufficient sensors as the representation for $f$ \cite{lu2021learning,lu2022comprehensive,meng2022learning}, which however will be computationally expensive for high-dimensional problems. To enhance the computational efficiency for the high-dimensional problem, we propose the following two alternatives: (1) we randomly select a certain number of measurements for $f$, and apply the PCA to them. The coefficients for the first $K$ components can then be used as the representation for $f$; and (2) we can train a DNN to learn the representation for $f$, which takes as input the paired data $(\bm{x}_i, f(\bm{x}_i))$ and outputs the representation of $f$,  as the summary network in \cite{radev2020bayesflow}. The latter approach is also suitable for the case where the numbers and/or the locations of the measurements are different for each snapshot as well as the case with unstructured data.  

\section{Results and discussion}
\label{sec:results}
In this section, we first employ sGeM to learn stochastic processes, which can be achieved by ignoring the physics-informed parts in Sec. \ref{sec:method}. Specifically, two Gaussian processes (GP) with different squared exponential kernels and a non-Gaussian process are considered. We then test forward and inverse SDE problems using sPI-GeM. To further demonstrate the scalability of the proposed approach, we also utilize the present method to solve an SDE with high-dimensional stochastic and spatial space, i.e., $D_{\bm{\zeta}} > 50$ and $D_{\bm{x}} = 20$. Details on the computations, e.g., architectures, training steps of sGeM/PI-sGeM, etc., for each case are presented in \ref{sec:computations}.

\subsection{sGeM for stochastic processes}
\label{sec:part_1}

% \subsubsection{GP: $\sigma_l=0.1$}
We aim to approximate Gaussian and non-Gaussian processes using sGeM given training data. Specifically, the training data are generated from the following stochastic processes:
\begin{equation}\label{eq:gp}
u \sim \mathcal{GP}(0, \kappa(x, x')), \kappa(x, x') = \exp[\frac{-(x - x')^2}{2l^2}],~x, x' \in [-1, 1], 
\end{equation}
and 
\begin{equation}\label{eq:non-gp}
u \sim \exp[\mathcal{GP}(0, \kappa(x, x'))], ~\kappa(x, x') = \exp[\frac{-(x - x')^2}{2l^2}],~x, x' \in [-1, 1], 
\end{equation}
respectively. In Eqs. \eqref{eq:gp} and \eqref{eq:non-gp}, $\mathcal{GP}(0, \kappa(x, x'))$ denotes a Gaussian process with zero mean and the covariance function $\kappa(x, x')$, and $l$ is the correlation length.

\begin{figure}[H]
    \centering
    \subfigure[]{\label{fig:process_a}
    \includegraphics[width=0.36\textwidth]{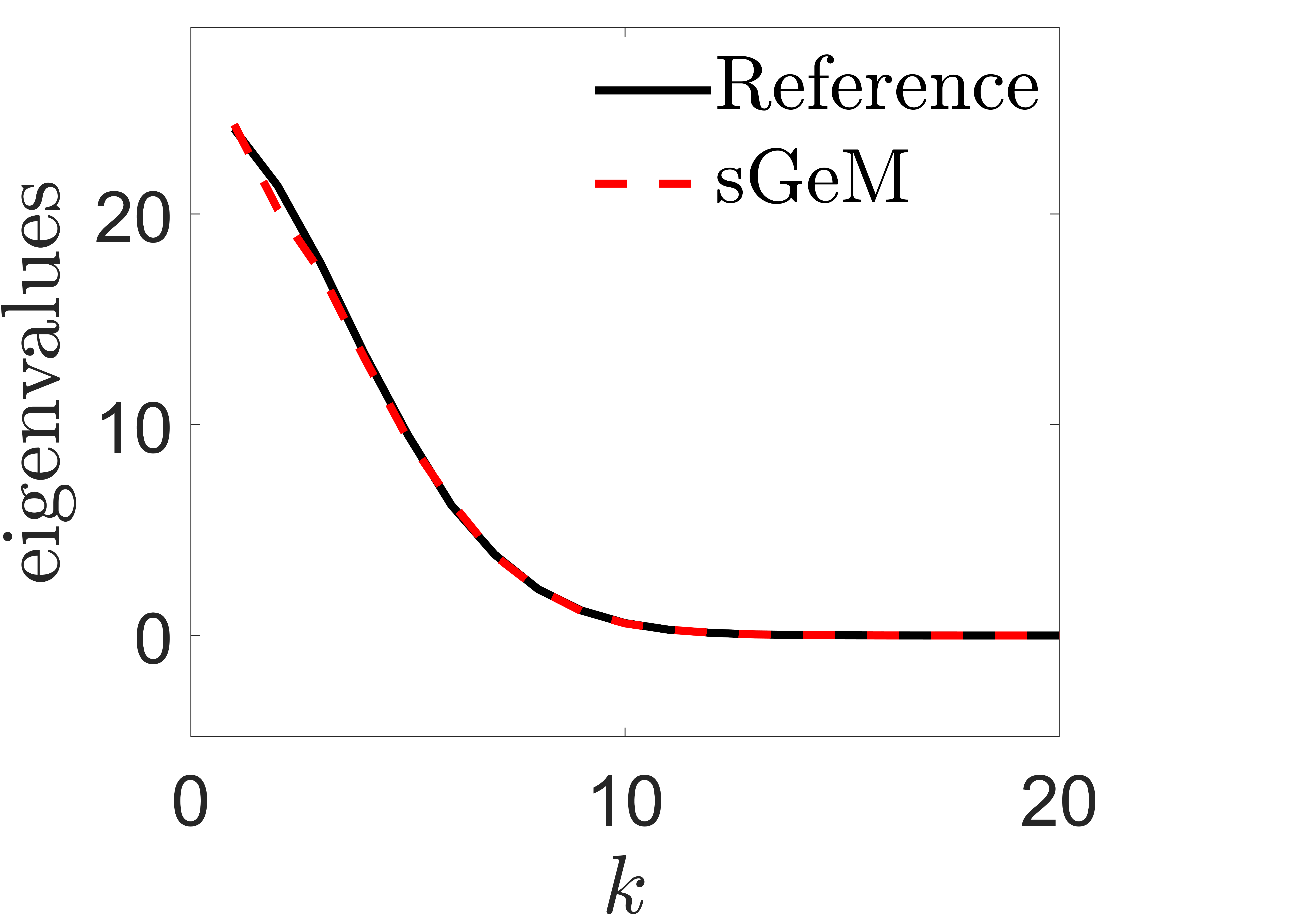}}\hspace{-1.cm}
    \subfigure[]{\label{fig:process_b}
    \includegraphics[width=0.36\textwidth]{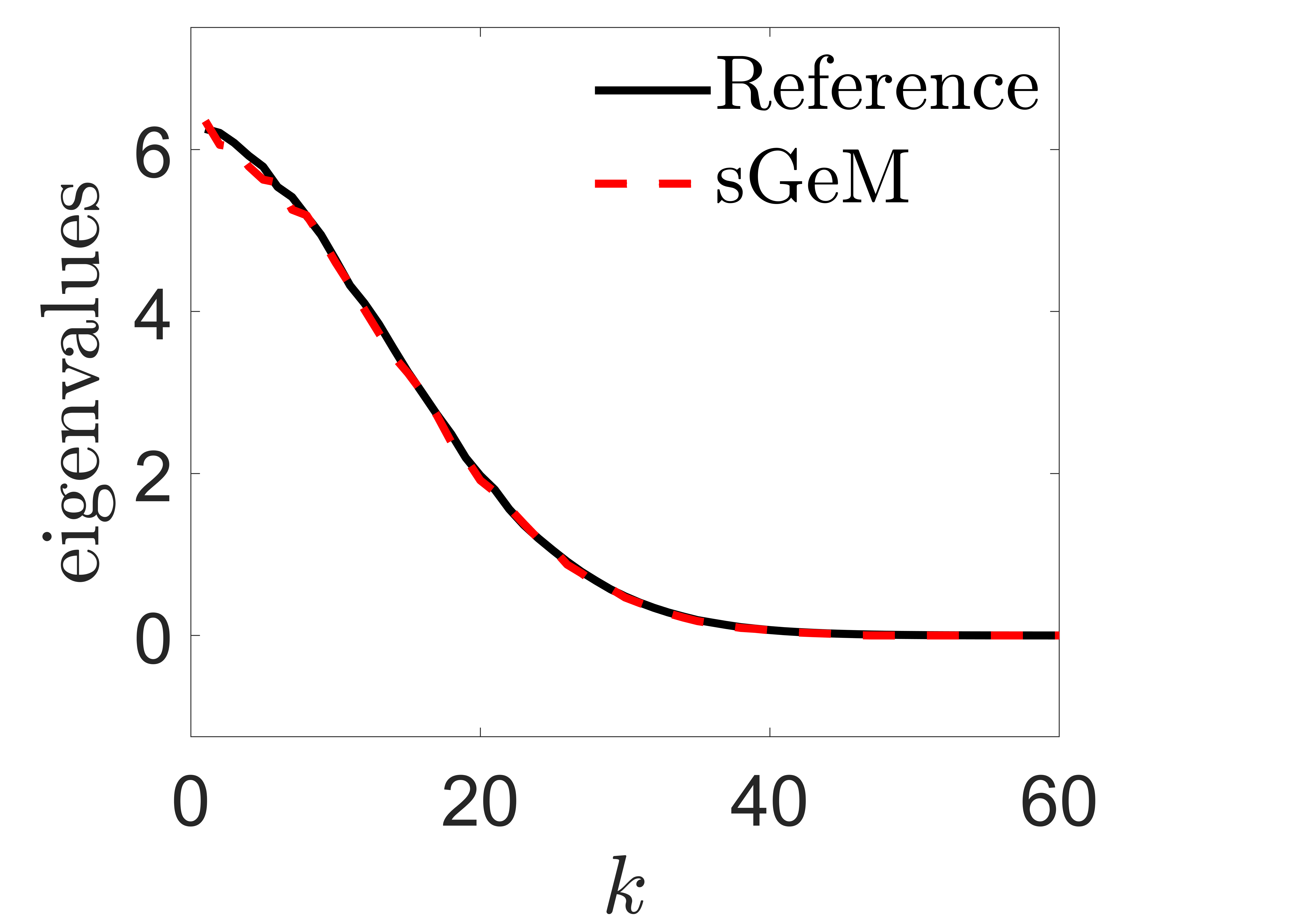}}\hspace{-1.cm}
    \subfigure[]{\label{fig:process_c}
    \includegraphics[width=0.36\textwidth]{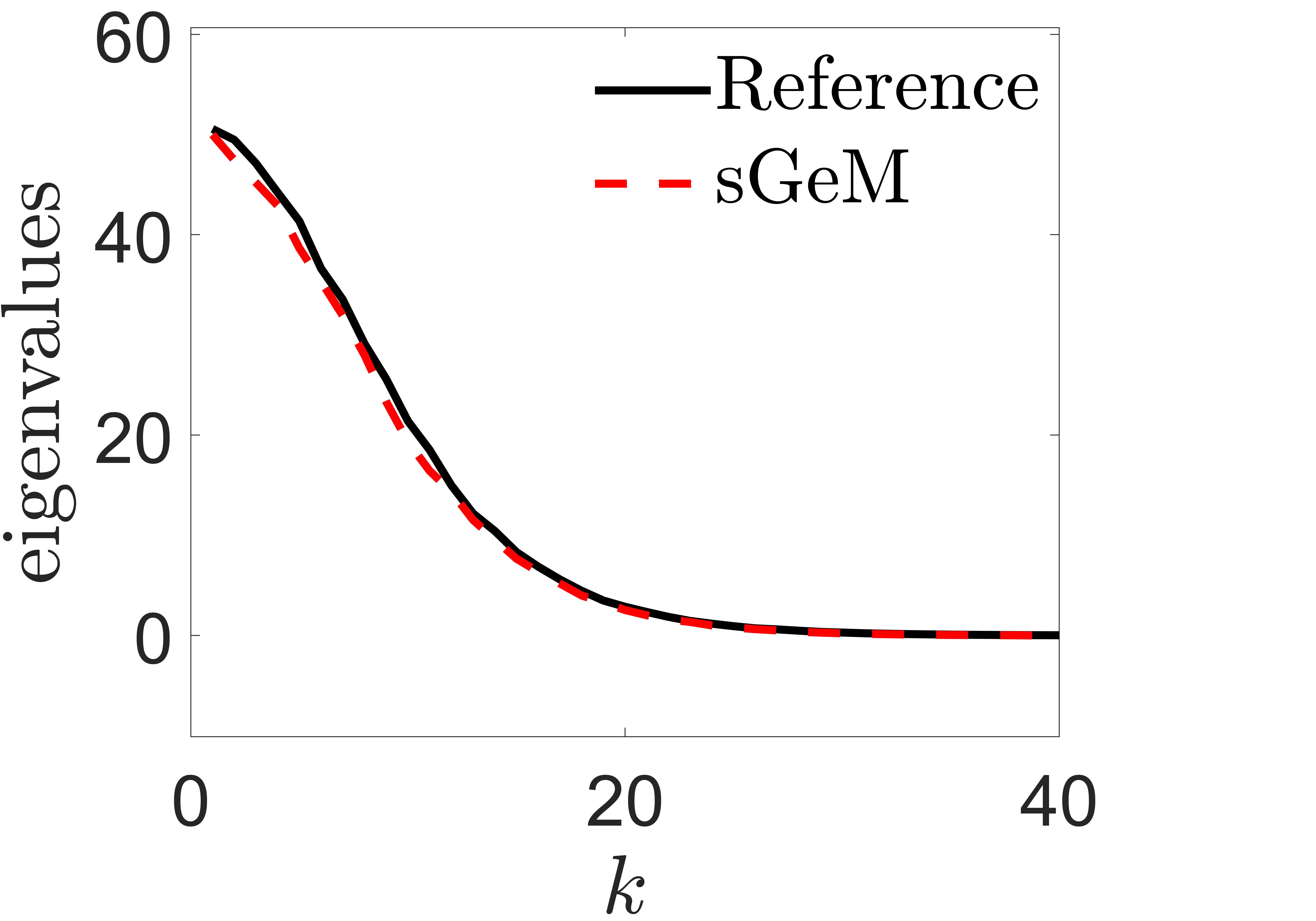}}\hspace{-1.cm}
    \caption{\label{fig:process}
    sGeM for approximating stochastic processes: Eigenvalues of the covariance matrix. (a) GP with  $l = 0.2$, (b) GP with  $l = 0.05$, and (c) a non-Gaussian process with $l = 0.1$. 
    }
\end{figure}

We first employ the sGeM to approximate the Gaussian process in Eq. \eqref{eq:gp}. In particular, Gaussian processes with two different correlation lengths are considered, i.e. $l = 0.2$ and $0.05$. For each test case, we randomly draw 10,000 snapshots or samples from the corresponding stochastic process as the training data. Each snapshot is resolved by 100 uniform measurements here. For this specific case, the input for $\mathcal{NN}_C$ is also the 100 equidistant measurements of $u$.  With the trained BasisNet, we can then learn the distribution of the coefficients for $\mathcal{D}_u$ using the sGeM. To justify the accuracy of the proposed approach, we generate 40,000 samples for $u$ using the trained GeM, and we compute the eigenvalues of the covariance matrix for each case based on the generated samples. In particular, we employ the sGeM to predict $u$ at 100 discrete points in each sample, and the points are equidistantly distributed for $x \in [-1, 1]$. As illustrated in Figs. \ref{fig:process_a} and \ref{fig:process_b}, the results from sGeM agree well with the reference solution for both cases, demonstrating the good accuracy of sGeM for approximating Gaussian processes. We further test the sGeM for approximating the non-Gaussian process, i.e. Eq. \eqref{eq:non-gp}, where $l = 0.1$. The setup for the training data is kept the same as in the previous two cases. Similarly, we present the eigenvalues for the covariance matrix obtained using 40,000 samples generated from the trained sGeM in Fig. \ref{fig:process_c}, which again shows the good accuracy of the proposed method. We note that: (1) the covariance functions in the Gaussian processes serve as the reference solutions in the first two cases, and (2)  the reference solution is obtained by computing the eigenvalues of the covariance matrix for 40,000 samples drawn from the non-Gaussian process since there is no exact solution for this particular case.  Effects of the number of measurements for each snapshot in the training data, and the number of snapshots on the computational accuracy are tested based on the non-Gaussian process. Interested readers are directed to  \ref{sec:converge_study}.

\begin{figure}[H]
    \centering
    \subfigure[]{\label{fig:process_loss_a}
    \includegraphics[width=0.36\textwidth]{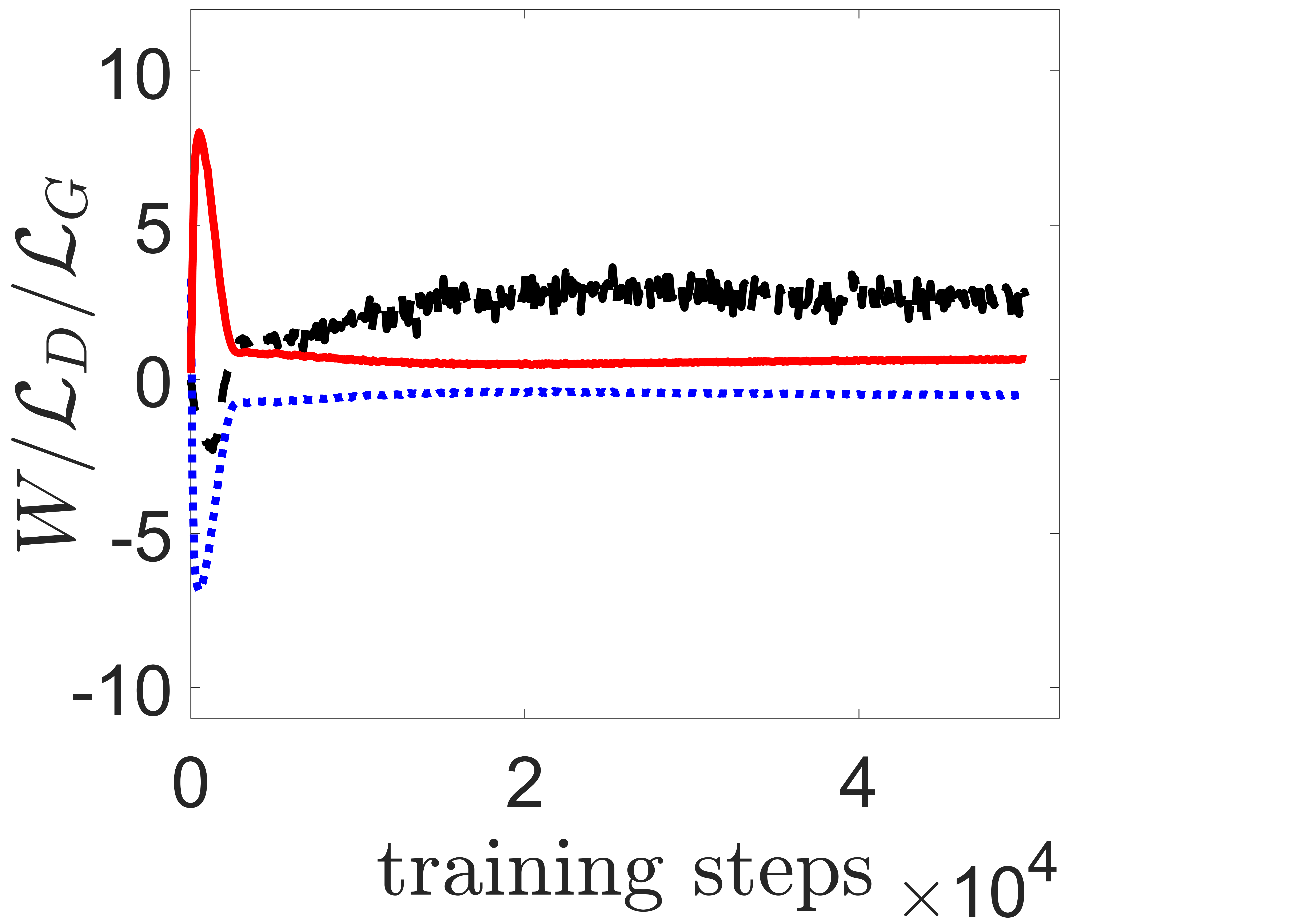}}\hspace{-1.cm}
    \subfigure[]{\label{fig:process_loss_b}
    \includegraphics[width=0.36\textwidth]{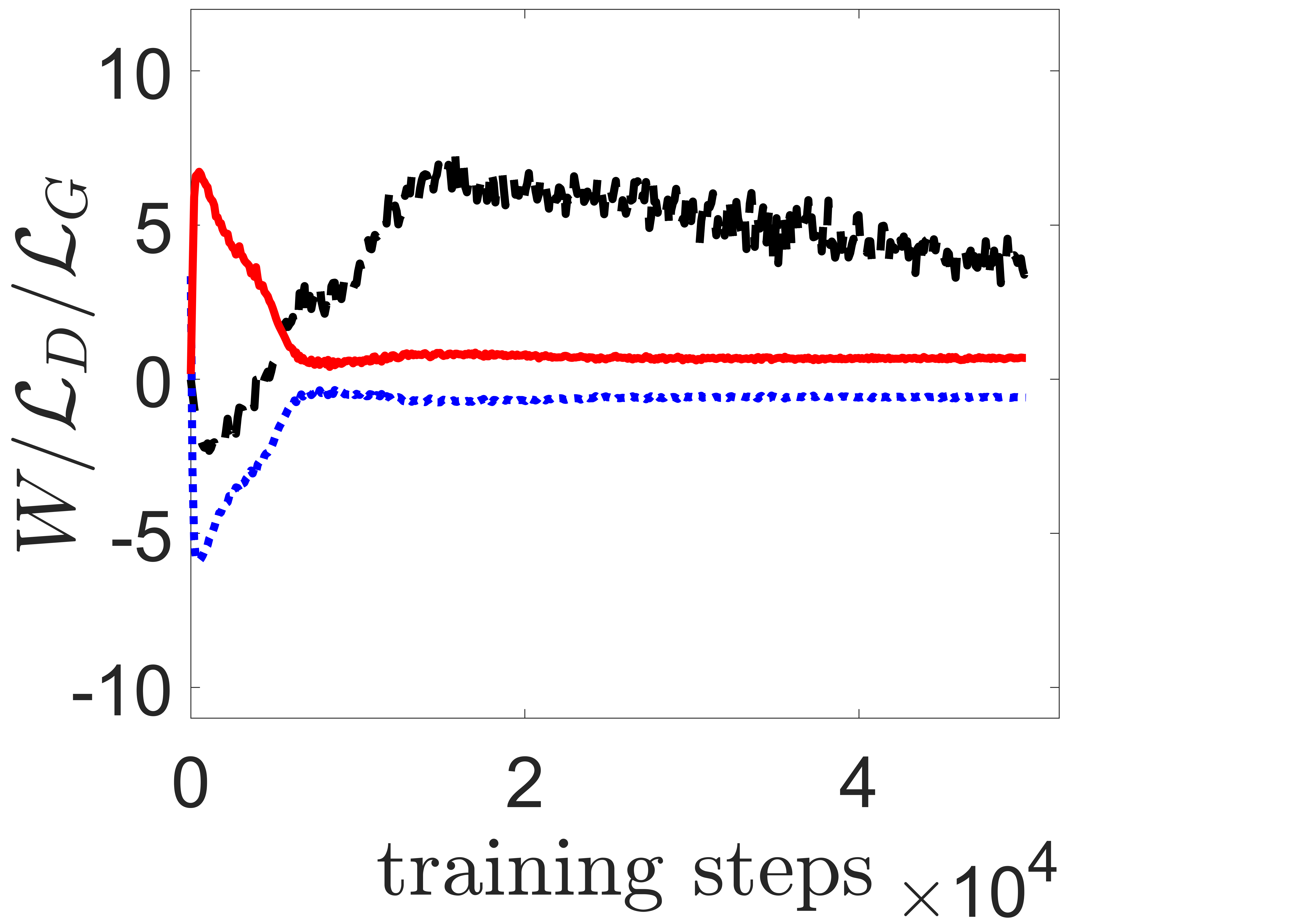}}\hspace{-1.cm}
    \subfigure[]{\label{fig:process_loss_c}
    \includegraphics[width=0.36\textwidth]{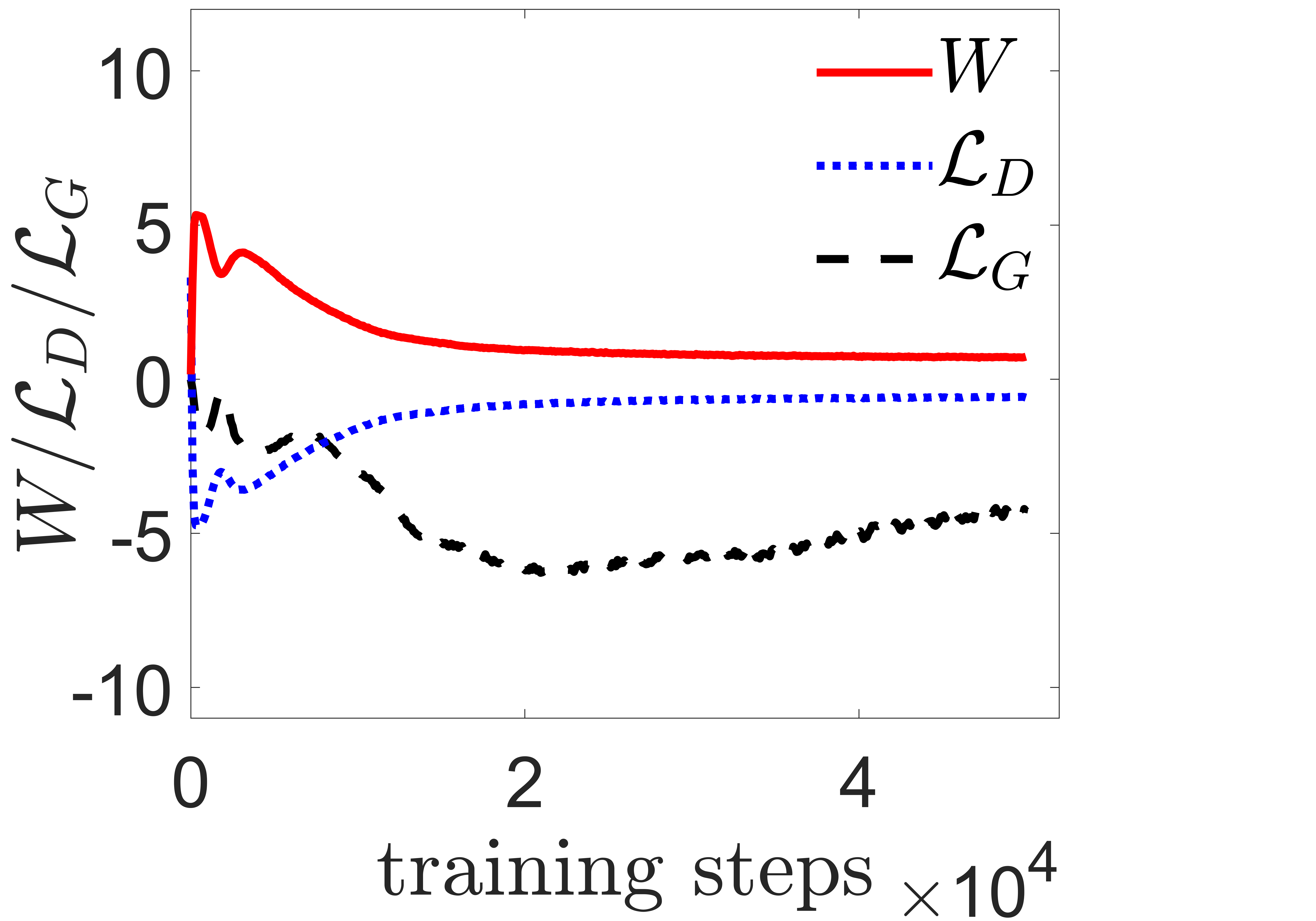}}\hspace{-1.cm}
    \caption{\label{fig:process_loss}
    sGeM for approximating stochastic process: Loss history for (a) GP with  $l = 0.2$, (b) GP with  $l = 0.05$, and (c) a non-Gaussian process with $l = 0.1$.
    $\mathcal{L}_D$: Loss for the discriminator; $\mathcal{L}_G$: Loss for the generator; $\bm{W}$: Estimation of the Wasserstein-1 distance.}
\end{figure}

\begin{figure}[H]
    \centering
    \subfigure[]{\label{fig:compare_step}
    \includegraphics[width=0.45\textwidth]{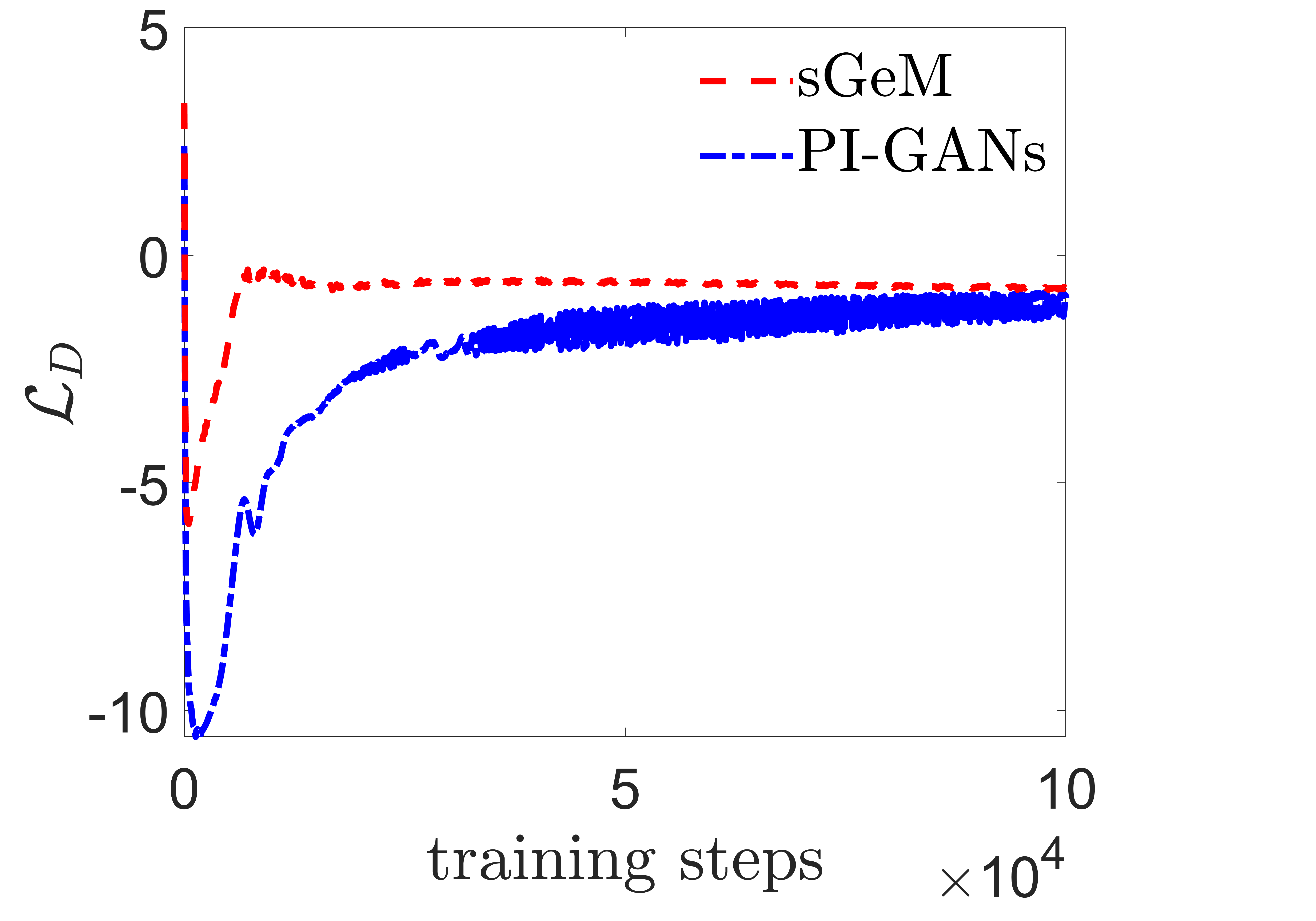}}
    \subfigure[]{\label{fig:compare_time}
    \includegraphics[width=0.45\textwidth]{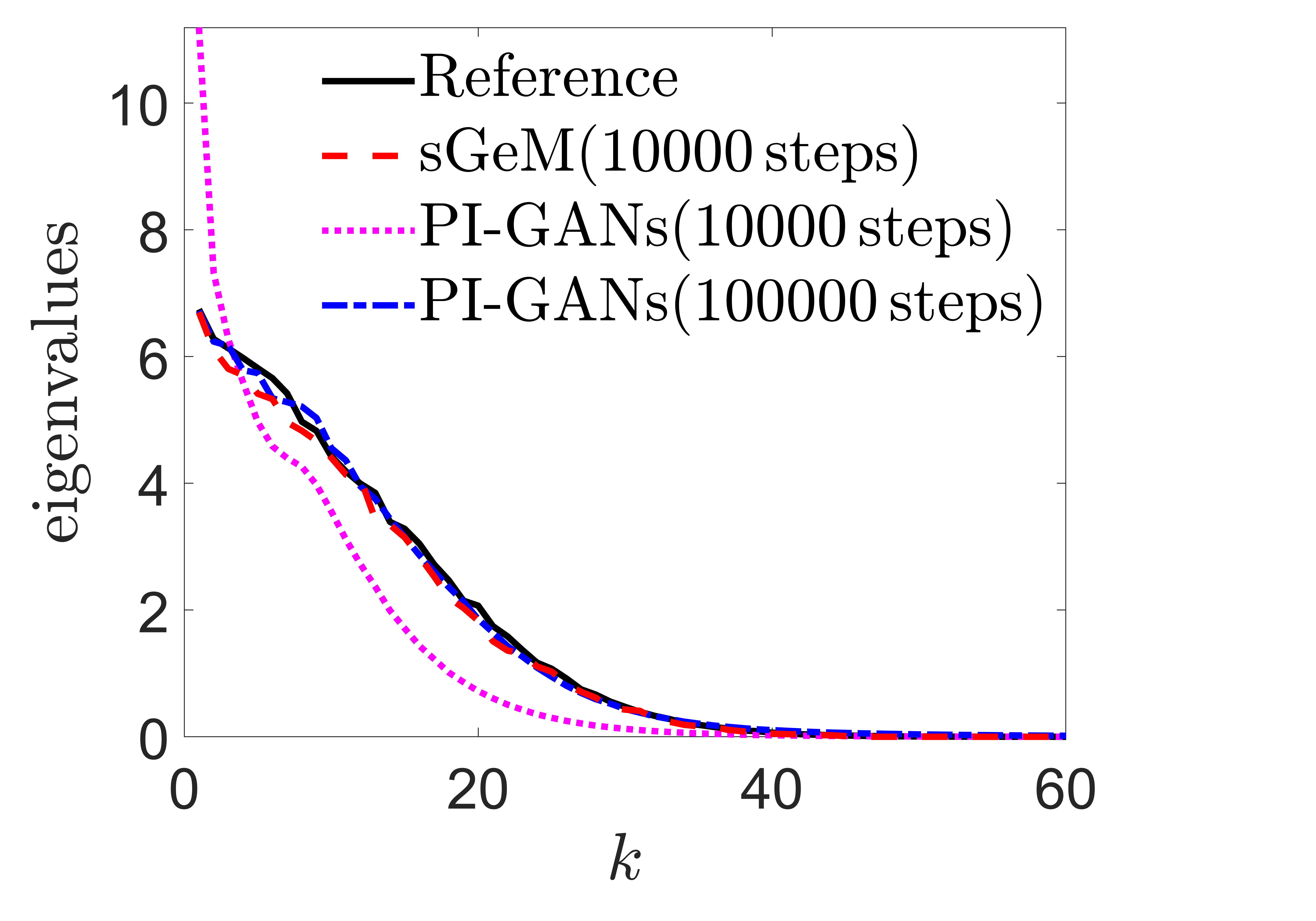}}
    \caption{\label{fig:compare} sGeM for approximating GP with $l = 0.05$: (a) loss histories for sGeM and PI-GANs \cite{yang2020physics}; (b) eigenvalues of the covariance matrix from sGeM and PI-GANs \cite{yang2020physics}.}
\end{figure}

{
To demonstrate the convergence of sGeM, we depict the loss histories of sGeM in the above test cases in Fig. \ref{fig:process_loss}. It is observed that the losses for the discriminator saturates in 10,000 and 25,000 for the Gaussian and non-Gaussian processes, respectively. Further, we conduct a comparison on the computational efficiency of sGeM and the PI-GAN developed in \cite{yang2020physics} based on the example in Fig. \ref{fig:process_b}. As shown in Fig. \ref{fig:compare}, we can see that the numbers of training steps required for convergence of sGeM and PI-GAN are about 10,000 and 100,000, respectively, which is expected as discussed in Sec. \ref{sec:method}. Furthermore, the computational time for the sGeM (including the training of BasisNet and GANs) and PI-GAN in this specific case are about 442 and 5208 seconds, respectively. The above results demonstrate that the sGeM is computationally more efficient compared to the PI-GANs for the specific case considered here. Note that the architectures of sGeM and PI-GANs, the parameters in the optimizer, etc., are kept the same in these two methods. Also, all the computations are performed on one NVIDIA GeForce RTX 3090. More details on the computations are directed to \ref{sec:computations}.
}

% As mentioned in Sec. \ref{sec:method}, the present method is expected to achieve faster convergence comparing to the approaches proposed in \cite{yang2020physics,zhong2023pi}. 
% We then depict the loss histories of sGeM in the above test cases in Fig. \ref{fig:process_loss}. It is observed that the losses for the discriminator saturates in 10,000 and 25,000 for the Gaussian and non-Gaussian processes, respectively. For comparison, it takes more than 100,000 training steps to convergence as reported in \cite{yang2020physics,zhong2023pi} for their developed methods to approximate the stochastic process. 
% \textcolor{red}{In Fig.\ref{fig:compare} we compared the training efficiency of sGeM and PI-GANs\cite{yang2020physics} using identical hardware configurations, training datasets, and optimizer as well as hyperparameters in the optimizer, differing only in network architecture. To evaluate the generative quality during training, we sampled 40,000 instances to compute their eigenvalues and calculated the Mean Squared Error against reference values. The results demonstrate that our model exhibits a faster convergence rate for a given number of iterations. Furthermore, it requires significantly less computational time to achieve the same level of accuracy.}

Finally, the metric used to measure the dissimilarity between two stochastic processes in the present model is the Wasserstein-1 ($\bm{W}-1$) distance, which can be estimated by the discriminator of GeM. As shown in Fig. \ref{fig:process_loss}, the $\bm{W}-1$ distances share the same trend with the losses of the discriminator in each case, and will not be discussed in detail here. Note that we employ $\bm{W}$ to denote the $\bm{W}-1$ distance in all the figures of this study for simplicity.

\subsection{Forward stochastic Helmholtz  problem ($D_{\bm{\zeta}} = 52$ and $D_{\bm{x}} = 2$)}
\label{sec:high_d_zeta}

% Requires: \usepackage{amsmath}

We now test the sPI-GeM for solving forward SDE problems. Specifically,  a stochastic Helmholtz equation with two dimensions in spatial domain and 52 dimensions in the stochastic domain is considered, which is expressed as:
\begin{equation}
    -\Delta u(\bm{x},\bm{\zeta}) + \lambda u(\bm{x},\bm{\zeta}) = f(\bm{x},\bm{\zeta}), \quad \bm{x} = (x, y), ~ x, y \in [-\pi, \pi],
    \label{eq:Helmholtz}
\end{equation}
where $u(\bm{x}, \bm{\zeta})$ denotes the solution, $\Delta$ is the Laplacian operator, $\lambda = 1$ is a known constant, and $f(\bm{x}, \bm{\zeta})$ is a stochastic forcing term. We would like to obtain the solution $u$ given snapshots for $f(\bm{x}, \bm{\zeta})$ as well as the boundary conditions. In particular, the exact solution of $u$ considered here reads as
\begin{equation}\label{eq:sde_50_zeta}
\begin{aligned}
    u(\bm{x}, \bm{\zeta})=\frac{9}{100}&[(x^2-\pi^2)\sum_{n=1}^{d/4}\frac{1}{n^2}(\zeta_{n} \text{sin}(nx)+\zeta_{n+\frac{d}{4}} \text{cos}(nx)) \times \\
     & (y^2-\pi^2)\sum_{n=1}^{d/4}\frac{1}{n^2}(\zeta_{n+\frac{2d}{4}} \text{sin}(ny) +\zeta_{n+\frac{3d}{4}} \text{cos}(ny))],
\end{aligned}
\end{equation}
where $d = 52$, and $\zeta_i \sim U(0, 1), i = 1, ..., d$. The boundary conditions and $f(\bm{x}, \bm{\zeta})$ can then be  obtained accordingly given Eq. \eqref{eq:sde_50_zeta}.

Similarly, we first train the PI-BasisNet to obtain the basis functions as well as the corresponding coefficients for $u$. For this test case, we first randomly draw 5,000 samples of $u$ from Eq. \eqref{eq:sde_50_zeta}, and compute the corresponding $f$ to serve as the training data. The input for $\mathcal{NN}_C$ of PI-BasisNet, i.e.,  \(\mathcal{U}\), is represented by the measurements for $f$ on a $x \times y = 128 \times 128$ uniform grid. In addition, we employ a convolutional neural network (CNN) for $\mathcal{NN}_C$ in the PI-BasisNet to achieve good computational efficiency as in \cite{lu2022comprehensive, xiao2024neural}. The boundary conditions are hard encoded to $\mathcal{NN}_B$. The output of PI-BasisNet is the approximation to $u$, and Eq. \eqref{eq:Helmholtz} is then encoded in BasisNet to obtain the approximation to $f$.

\begin{figure}[H]
    \centering
    \subfigure[Predicted mean for $u$]{\label{fig:Helmholtz_mean}
    \includegraphics[width=0.48\textwidth]{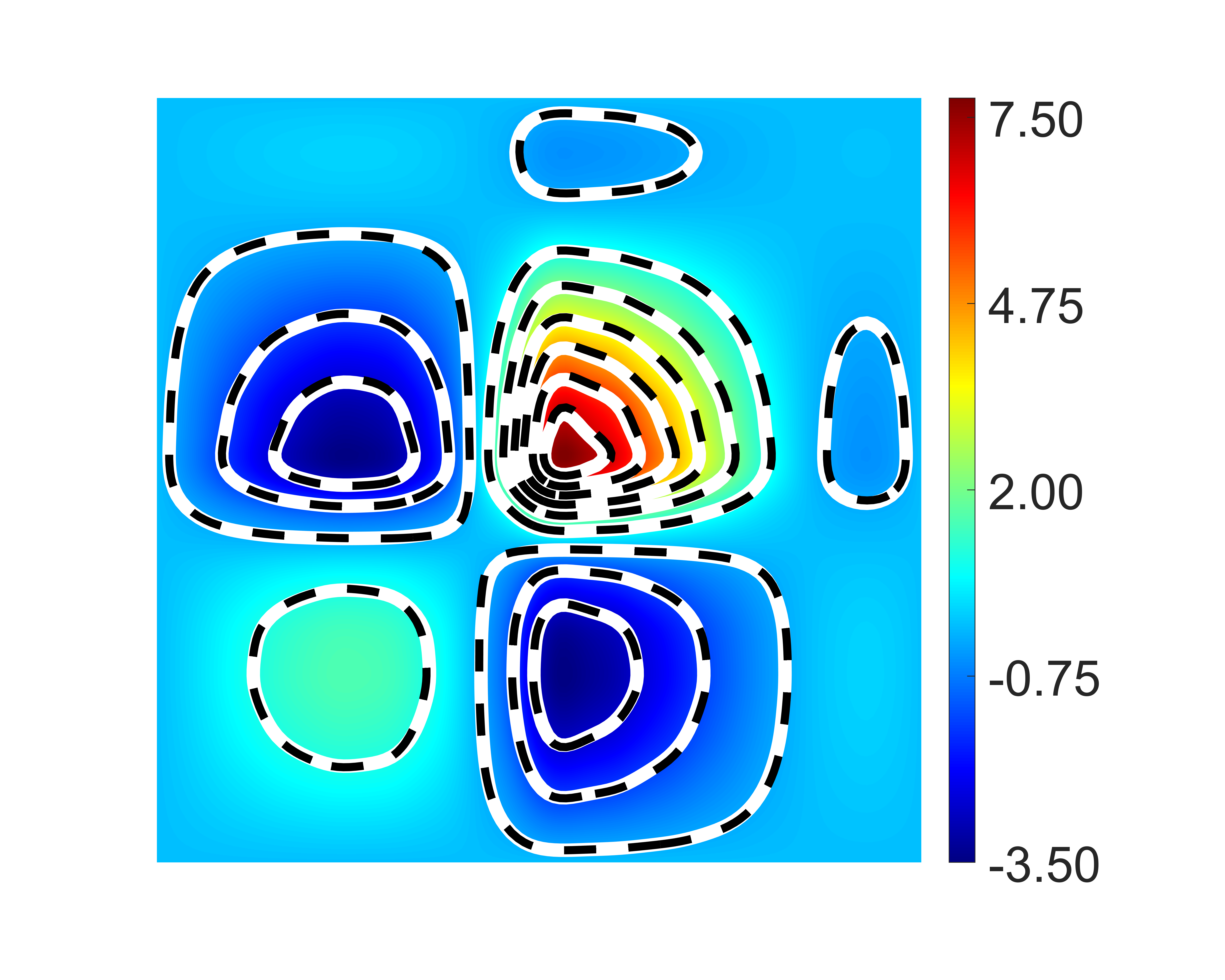}}\hspace{-8pt}
    \subfigure[Predicted standard deviation for $u$]{\label{fig:Helmholtz_std}
    \includegraphics[width=0.48\textwidth]{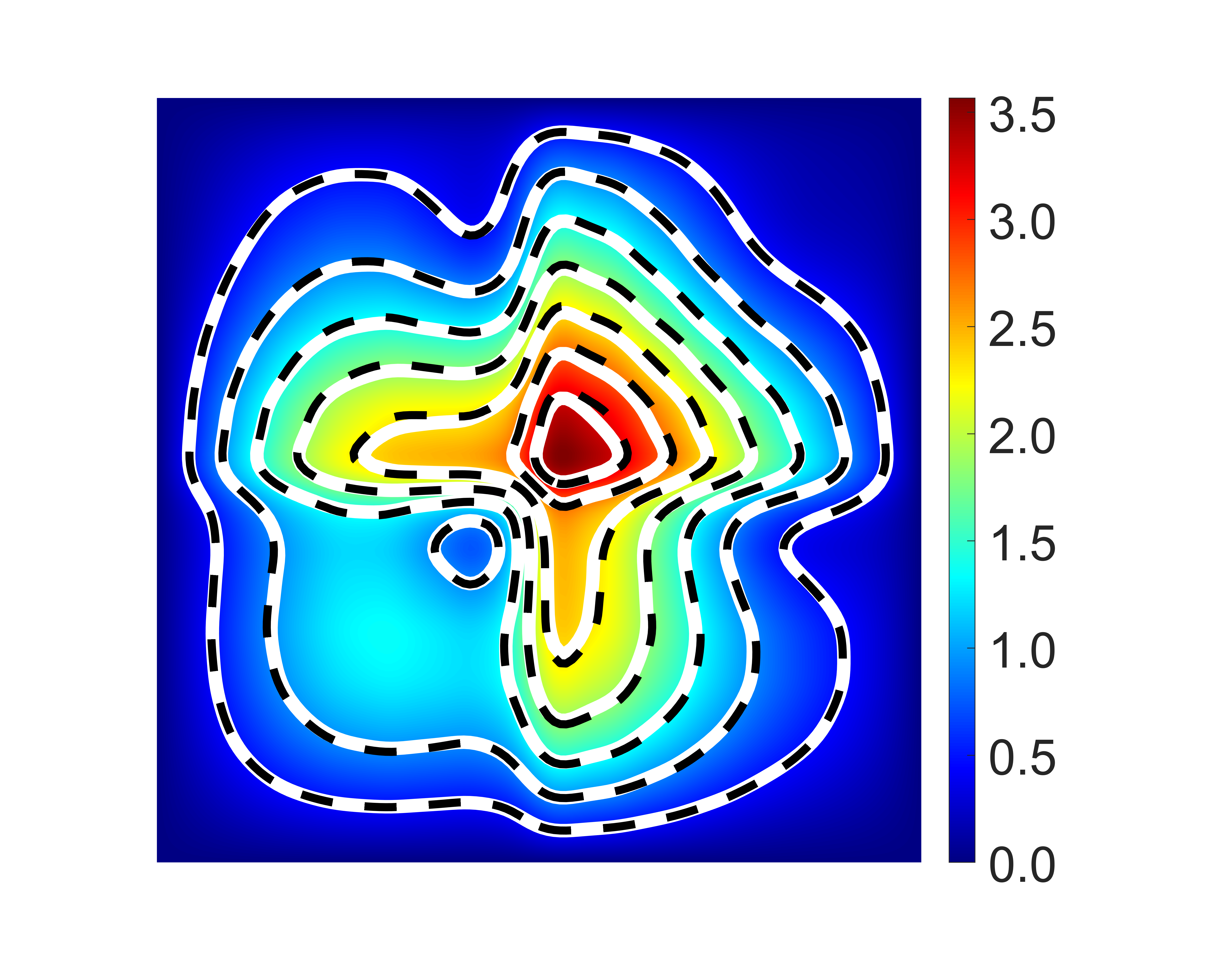}}
    
    \caption{\label{fig:Helmholtz_mean_std}
    sPI-GeM for forward stochastic Helmholtz  problem ($D_{\bm{\zeta}} = 52$ and $D_{\bm{x}} = 2$): Predicted (a) mean,  and (b) standard deviation for $u$. Colored background with white solid line: reference solution; Black dashed line: Predictions from sPI-GeM.}
\end{figure}

\begin{figure}[H]
    \centering
    \includegraphics[width=0.6\textwidth]{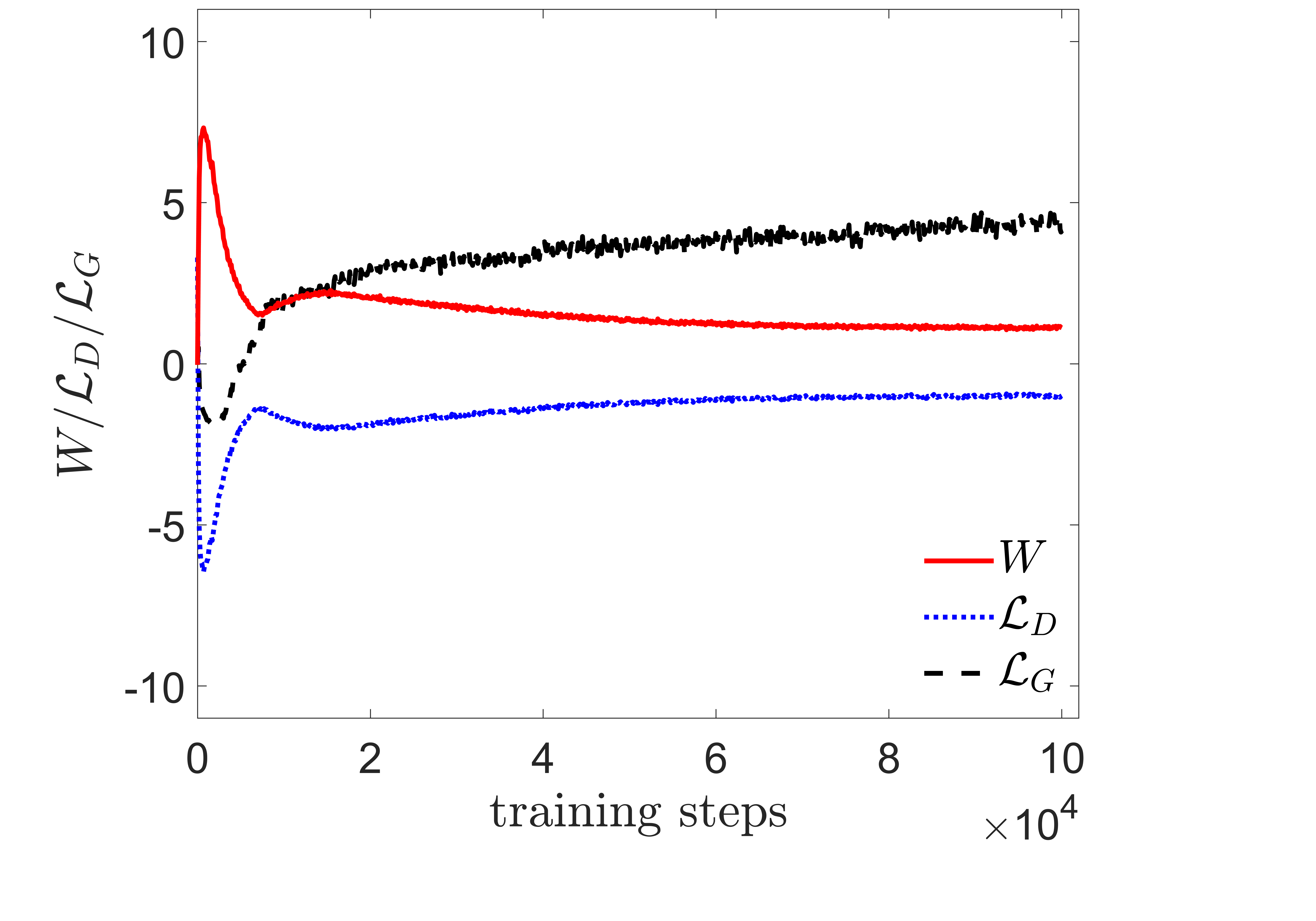}
    \caption{sPI-GeM for forward stochastic Helmholtz  problem ($D_{\bm{\zeta}} = 52$ and $D_{\bm{x}} = 2$): Loss history for sPI-GeM. $\mathcal{L}_D$: Loss for the discriminator; $\mathcal{L}_G$: Loss for the generator; $\bm{W}$: Estimation of the Wasserstein-1 distance.}
    \label{fig:Helmholtz_std_loss}
\end{figure}

With the trained PI-BasisNet, we can obtain the coefficients for $u$ related to the given data on $f$, which is then utilized as the training data for learning the distribution over the coefficients for $u$ with the PI-GeM.  Further, we can generate new samples for $u$ and $f$ upon the training of PI-GeM following Eq. \eqref{eq:output_sgem}. Specifically, 
we employ the trained sPI-GeM to generate 10,000 samples for \(u\) to compute the corresponding mean and standard deviation (std). In addition, each sample is resolved by a $128 \times 128$ uniform grid. As illustrated in Fig. \ref{fig:Helmholtz_mean_std}, both the predicted mean and standard deviation show little discrepancy compared to the reference solution. Here the reference is obtained using 10,000 samples generated from the exact solution in Eq. \eqref{eq:sde_50_zeta}.  We further present the computational errors for $u$ and $f$ between the predictions from the sPI-GeM and the reference solutions in Table \ref{tab:Helmholtz_tab}. The above results demonstrate the capability of sPI-GeM to handle SDE problems with high-dimensional stochastic space. Also, we illustrate the loss history of the sPI-GeM in Fig. \ref{fig:Helmholtz_std_loss}, it can be seen that the loss for the discriminator is quite smooth and the present approach converges in around 30,000 training steps for this particular case. 

As reported in \cite{rahaman2019spectral,xu2024overview}, it is challenging for DNNs to fit functions with high frequency well due to the spectral bias. The feature expansion is an effective approach to address this issue \cite{lu2022comprehensive}. Here we also enhance the $\mathcal{NN}_{B}$ with feature expansion in this test case, and the accuracy can be further improved as illustrated in Table \ref{tab:Helmholtz_tab}. Details on the feature expansion used here are present in \ref{sec:computations}.

\begin{table}[H]
\centering
\caption{sPI-GeM for forward stochastic Helmholtz  problem ($D_{\bm{\zeta}} = 52$ and $D_{\bm{x}} = 2$):  Relative $L_2$ errors for $u$ and $f$.}
\begin{tabular}{lcccc}
\toprule
& \textbf{\thead{predict mean\\Rel. $L_2$ Error}} & \textbf{\thead{predict std \\Rel. $L_2$ Error}}\\
\midrule
$u$ (w/o feature expansion) & $4.42\%$  & $3.00\%$ \\
$f$ (w/o feature expansion) & $4.11\%$  & $3.28\%$ \\
\hline
$u$ (w/ feature expansion) & $1.64\%$  & $2.85\%$ \\
$f$ (w/ feature expansion) & $3.58\%$  & $3.20\%$ \\
\bottomrule
\label{tab:Helmholtz_tab}
\end{tabular}
\end{table}

{
\subsection{2D Darcy flow in heterogeneous porous media}
\label{sec:darcy_2d}
% In this section, we investigate the stochastic Darcy equation governing two-dimensional flow in heterogeneous porous media using the sPI-GeM, focusing on the forward propagation of uncertainty. Specifically, we analyze how stochastic variations in hydraulic conductivity influence the resulting hydraulic head distribution. The steady-state flow is described by \cite{zheng2020physics}:
In this section, we employ the sPI-GeM to solve  a two-dimensional stochastic Darcy flow problem in heterogeneous porous media, which is a widely used benchmark problem for solving  SDEs. The governing equation for this specific problem is described as follows \cite{zheng2020physics}:
\begin{equation}
    -\nabla \cdot \big(\lambda(\bm{x},\bm{\zeta}) \nabla u(\bm{x},\bm{\zeta}) \big) = 0, ~\bm{x}=(x,y),~ x,y \in [-1, 1],
    \label{eq:Darcy}
\end{equation}
with boundary conditions
\begin{equation}
\begin{aligned}
    u(x,-1)&=1,~ u(x,1)=0,\\ \partial_{\bm{n}}u(-1,y)&=0,~\partial_{\bm{n}}u(1,y)=0,
\end{aligned}
\end{equation}
where $\lambda$ represents the random hydraulic conductivity, $u$ denotes the hydraulic head, and $\bm{n}$ denotes the unit outward normal vector to the boundary. Specifically, $\lambda(\bm{x},\bm{\zeta})$ is expressed as  $\lambda(\bm{x},\bm{\zeta}) = \exp(0.5F(\bm{x},\bm{\zeta}))$ following \cite{zheng2020physics}. In particular, $F(\bm{x},\bm{\zeta})$ is defined as a zero-mean Gaussian process characterized by the following anisotropic squared-exponential  kernel:
\begin{equation}\label{eq:2d_gp}
\kappa(x, x') = \exp\left[\frac{-(x - x')^2}{2l_x^2}+\frac{-(y - y')^2}{2l_y^2}\right],~ x, x' \in [-1, 1],~ y, y' \in [-1, 1], 
\end{equation}
where $l_x=0.2$ and $l_y=0.4$ are the correlation lengths in the $x$ and $y$ directions, respectively. The objective is to solve Eq. \eqref{eq:Darcy} given multiple realizations of $\lambda$ as well as the boundary conditions. 

\begin{figure}[H]
    \centering
    \includegraphics[width=0.99\textwidth]{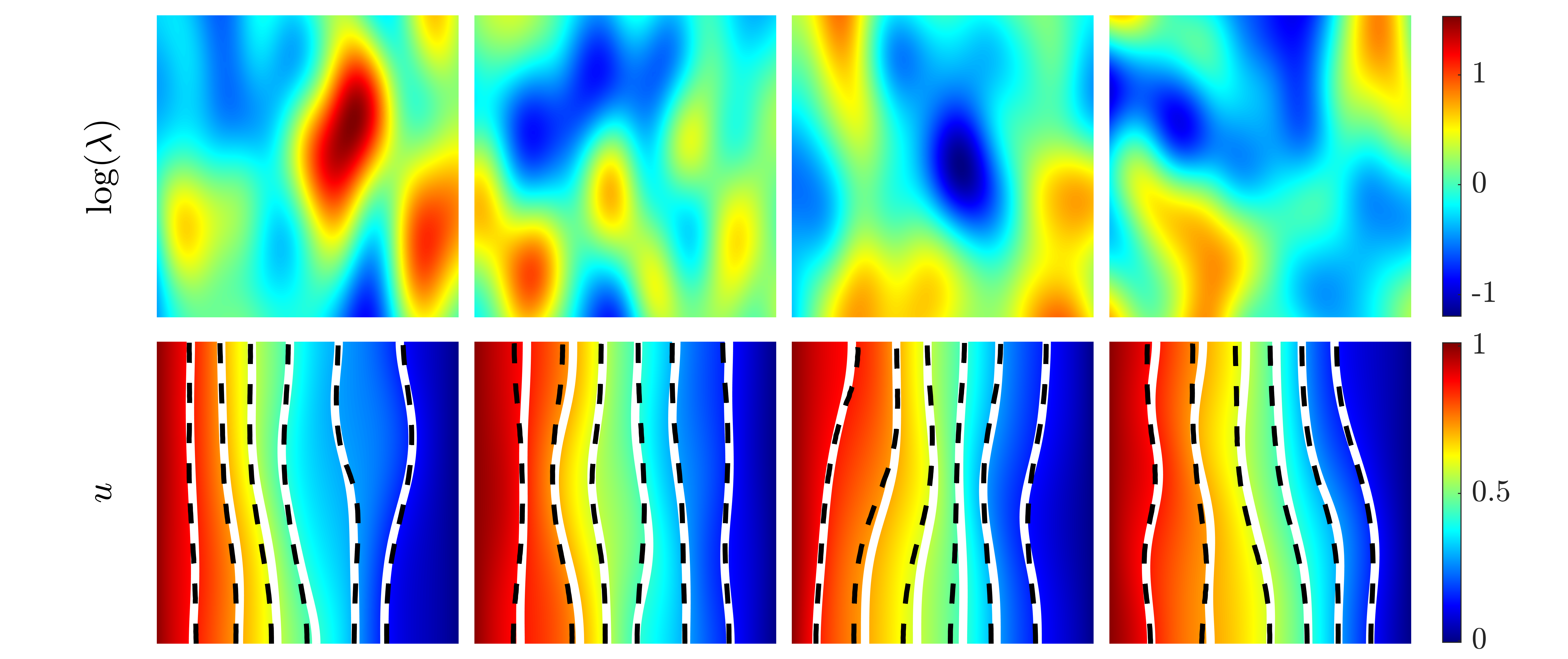}
    \caption{sPI-GeM for solving 2D Darcy flow in heterogeneous porous media: Predictions for the $\lambda$ and $u$. Colored background and white solid: reference solution; Black dashed line: predictions for $u$. Note that the reference solutions for $u$ are obtained from the {\sl Matlab PDE toolbox} given corresponding samples of $\lambda$ at the first row.}
    \label{fig:darcy}
\end{figure}

For this specific problem, we assume that we have 5,000 snapshots of $\lambda$, and each snapshot is resolved by a $101 \times 101$ uniform grid. Also, all the boundary conditions are known.  Similarly, we first train the PI-BasisNet to obtain the basis functions as well as the corresponding coefficients for $\lambda$ and $u$.  Specifically, we have one $\mathcal{NN}_C$ in the  PI-BasisNet to generate the coefficients for $\lambda$ and $u$, with the snapshot of $\lambda$ as input. Also,  a CNN is utilized for $\mathcal{NN}_C$ to achieve good computational efficiency. In addition, we use $\mathcal{NN}_{B,\lambda}$ and $\mathcal{NN}_{B,u}$ to learn the basis functions for $\lambda$ and $u$, respectively, as they have distinct features as shown in Fig. \ref{fig:darcy}.

% We first employ the BasisNet architecture  to encode the input field $K$, utilizing a dataset of 5,000 random fields sampled from a Gaussian Process. In addition, each sample is resolved by a $128 \times 128$ uniform grid. This stage serves purely to parameterize $K$ into its corresponding coefficients $\bm{\psi}_K$. Subsequently, we utilize a separate set of networks to represent the solution $h$ by $\bm{\psi}_h$. In this second stage, the network is trained by minimizing a physics-informed loss function Eq. \eqref{eq:Darcy} that depends on the hydraulic conductivity $K$, alongside the boundary conditions. Finally, we employ a GAN to learn the joint coefficient distribution of $K$ and $h$.

\begin{table}[ht]
\centering
\caption{sPI-GeM for 2D flow in heterogeneous porous media:  relative $L_2$ errors for  $u$.}
\begin{tabular}{lcccc}
\toprule
& \textbf{\thead{predict mean\\Rel. $L_2$ Error}}  & \textbf{\thead{predict std \\Rel. $L_2$ Error}}\\
\midrule
$u$ & $1.01\%$  & $4.55\%$ \\
% $\lambda$ & $2.85\%$  & $5.12\%$ \\
\bottomrule
\label{tab:darcy}
\end{tabular}
\end{table}

Similar as in the previous cases, we can obtain $\bm{\psi}_{\lambda}$ and $\bm{\psi}_u$ from the trained PI-BasisNet. We then train a sGeM to learn the distributions for $\bm{\psi}_{\lambda}$ and $\bm{\psi}_u$. Specifically, the input for the generative model is the sample from a  Gaussian distribution, and the output is the paired $(\bm{\psi}_{\lambda}, \bm{\psi}_u)$. Upon the training of sPI-GeM, we can then obtain the  predictions for paired $(\lambda, u)$. To demonstrate the accuracy of the present model, we first generate 10,000 samples for $(\lambda, u)$ from the trained sPI-GeM,  
and we then use the {\sl Matlab PDE toolbox} to solve Eq. \eqref{eq:Darcy} given the same number of samples of $\lambda$ generated from Eq. \eqref{eq:2d_gp} to obtain the reference solutions of $u$. We present several realizations for $(\lambda, u)$ from the sPI-GeM  in Fig. \ref{fig:darcy}, and we also illustrate the computed errors between the predictions and the reference solution for $u$ in Table \ref{tab:darcy}.  All the results in Fig. \ref{fig:darcy} and Table \ref{tab:darcy} again confirm the good accuracy of the present method. We finally illustrate the loss history for sPI-GeM in Fig. \ref{fig:Darcy_loss}. As shown, the discriminator in sPI-GeM converges in around 20,000 training steps. Details on the numerical methods used here are present in \ref{sec:computations}.

\begin{figure}[H]
    \centering
    \includegraphics[width=0.6\textwidth]{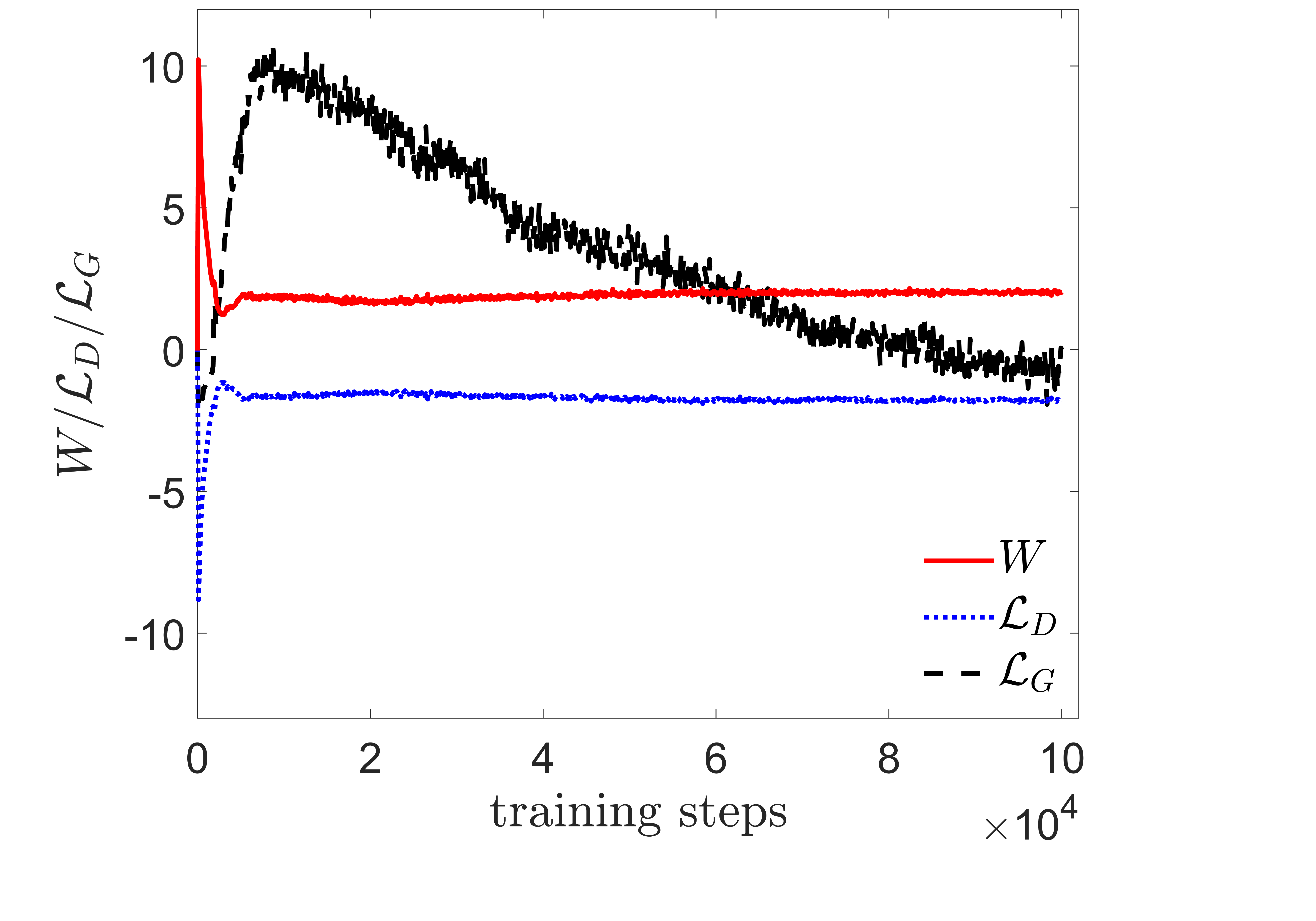}
    \caption{sPI-GeM for 2D flow in heterogeneous porous media: Loss history for sPI-GeM. $\mathcal{L}_D$: Loss for the discriminator; $\mathcal{L}_G$: Loss for the generator; $\bm{W}$: Estimation of the Wasserstein-1 distance.}
    \label{fig:Darcy_loss}
\end{figure}
}

\subsection{Inverse SDE problem}
\label{sec:inverse_pde}

We now consider to solve an inverse stochastic elliptic equation using the sPI-GeM. Note that the inverse problem considered here is the same as in \cite{yang2020physics,shin2023physics}. The equation is expressed as:

\begin{equation}    \label{eq:mixed_sde}
\begin{aligned}
    -\frac{1}{10} \frac{d}{dx} \left( \lambda(x,\bm{\zeta}) \frac{d}{dx} u(x,\bm{\zeta}) \right) &= f(x,\bm{\zeta}), ~ x \in [-1, 1],\\u(-1,\bm{\zeta}) &= u(1,\bm{\zeta})=0,
\end{aligned}
\end{equation}

where
% Requires: \usepackage{amsmath}
\begin{subequations}\label{eq:inverse_sde_f}
\begin{align}
    f(x, \bm{\zeta}) &\sim \mathcal{GP} \left( \frac{1}{2}, \frac{9}{400} \exp \left( -(x - x')^2 \right) \right), ~x, x' \in [-1, 1], \\
    \lambda(x; \bm{\zeta}) &= \exp \left( \frac{1}{5} \sin \left( \frac{3}{2} \pi (x + 1) \right) + \bar{\lambda} \right), ~x \in [-1, 1], \\
    \bar{\lambda} &\sim \mathcal{GP} \left( 0, \frac{1}{100} \exp \left( -(x - x')^2 \right) \right), ~x, x' \in [-1, 1].
\end{align}
\end{subequations}
In addition, $f(x, \bm{\zeta})$ and $\bar{\lambda}$ are Gaussian processes with squared exponential kernels. The diffusion coefficient $\lambda(x, \bm{\zeta})$ is thus a non-Gaussian stochastic process. We assume that we have partial observations on $u$ and $f$, and the objective is to obtain predictions for $u/f$ and $\lambda$ given data on $u$ and $f$.

In our computations, we assume that we have snapshots on $u$ and $f$. Specifically, we assume that (1) we have 10 sensors for $u$, which is uniformly distributed in $x \in [-0.95, 0.95]$, and (2) we have 100 sensors for $f$, which are uniformly distributed in $x \in [-1, 1]$, respectively. In addition, we assume that we have 5,000 snapshots for $u/f$ as training data. Similarly, we first use PI-BasisNet to learn the basis functions and the coefficients for $u$ and $\lambda$. The input for $\mathcal{NN}_C$ of the PI-BasisNet is $f/u$, which is represented by the measurements from the 100 sensors of $f$ and the 10 sensors of $u$. The output dimension for $\mathcal{NN}_C$ is 128, and the first half denotes the coefficients for $u$, while the remaining is used for the coefficients for $\lambda$.  In addition, we have two DNNs for learning the basis functions for $u$ and $\lambda$, respectively. The outputs of the PI-BasisNet are then approximations to $u$ and $\lambda$. By encoding Eq. \eqref{eq:mixed_sde} in the BasisNet, we can then obtain the approximation for $f$.

\begin{figure}[H]
    \centering
    \subfigure[]{\label{fig:1d_sde_inverse_u_a}
    \includegraphics[width=0.36\textwidth]{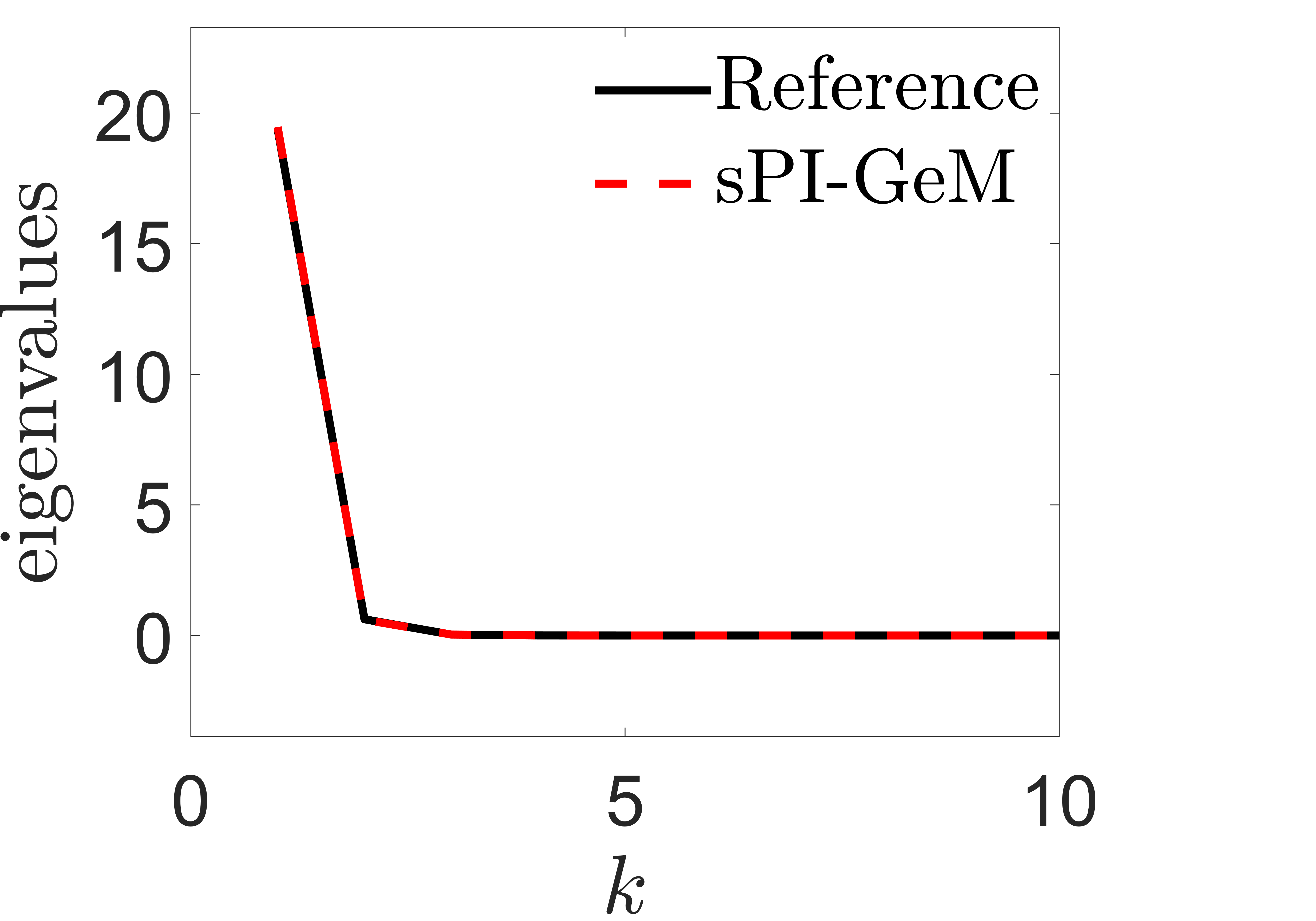}}\hspace{-1.cm}
    \subfigure[]{\label{fig:1d_sde_inverse_u_b}
    \includegraphics[width=0.36\textwidth]{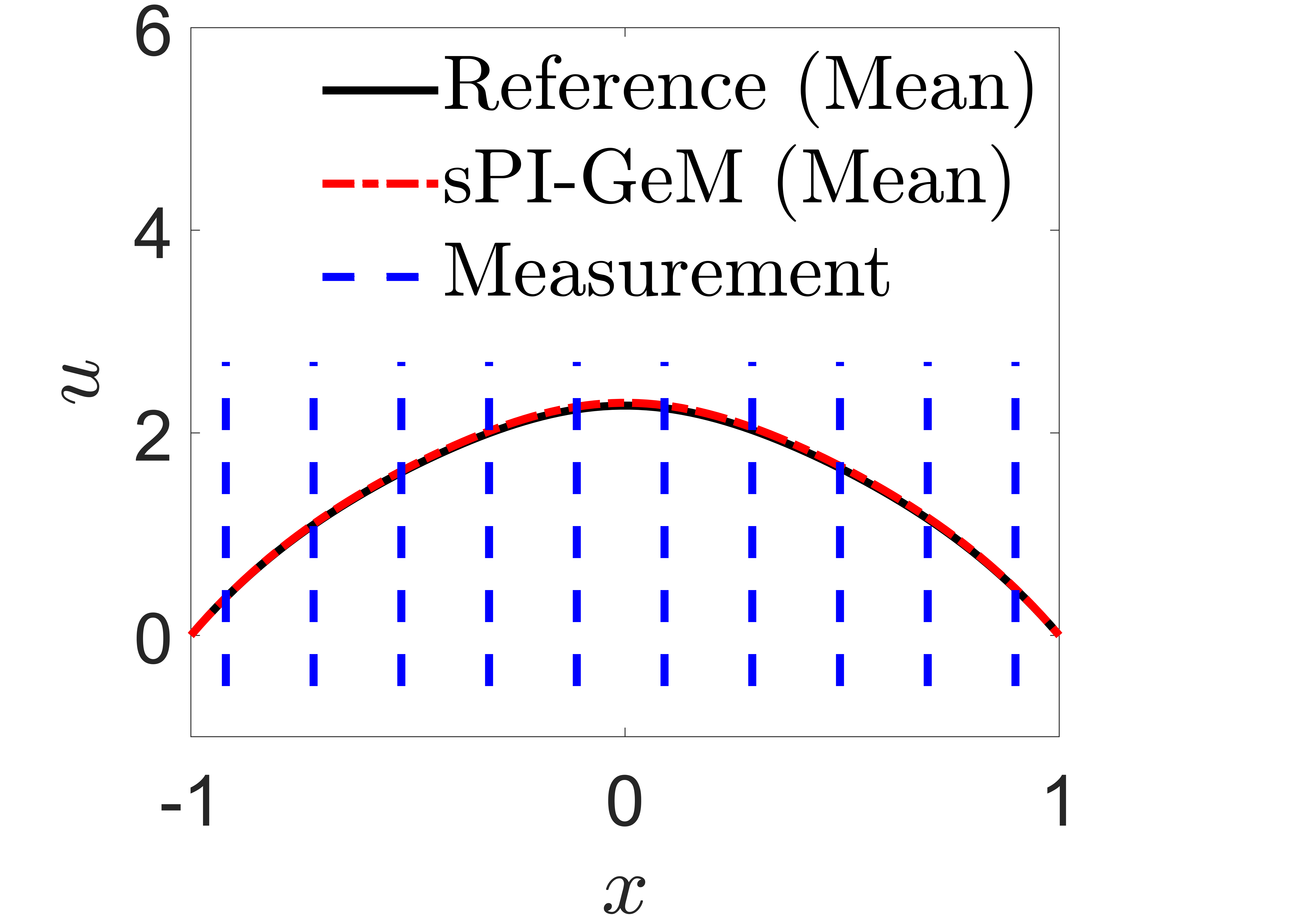}}\hspace{-1.cm}
    \subfigure[]{\label{fig:1d_sde_inverse_u_c}
    \includegraphics[width=0.36\textwidth]{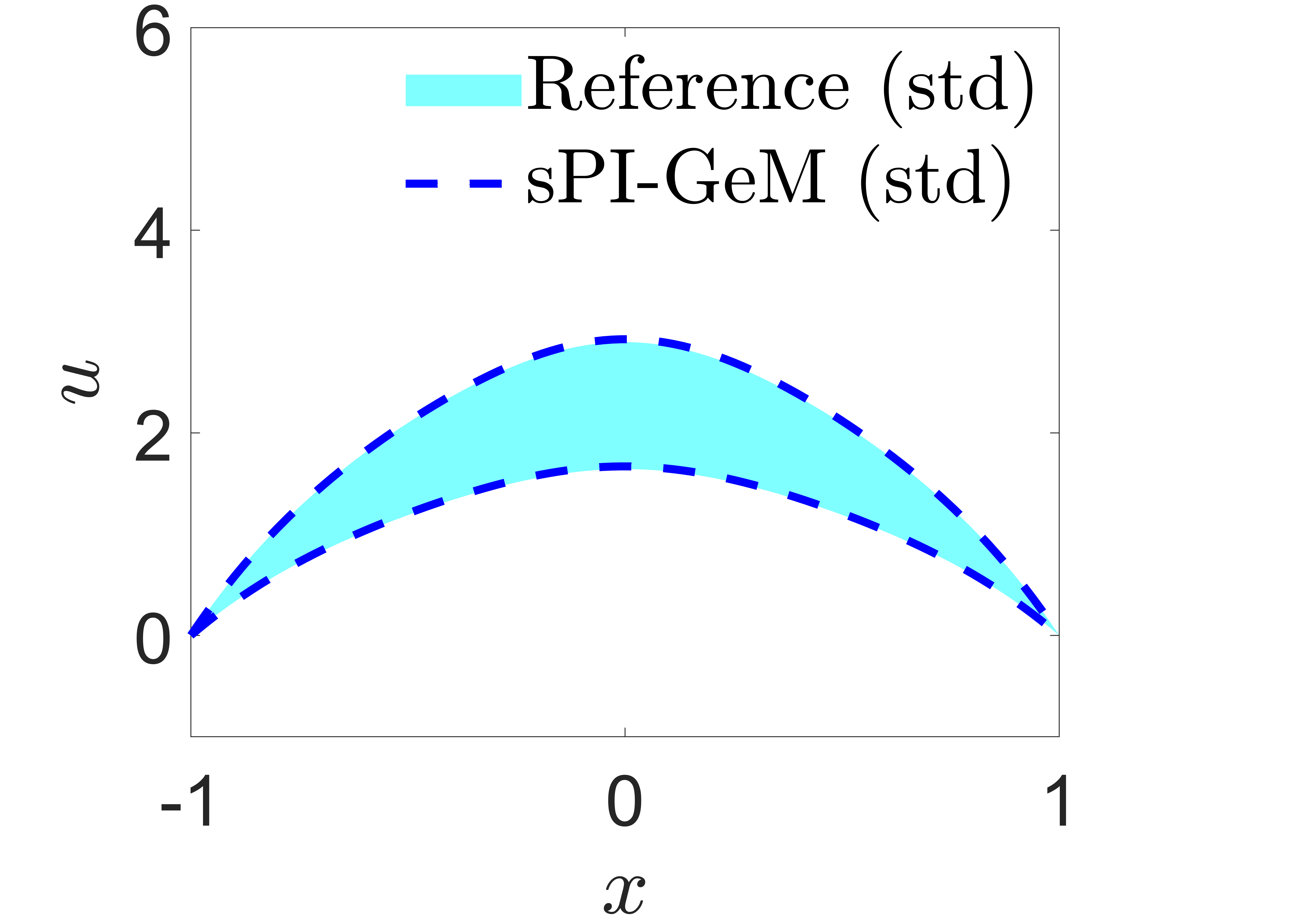}}\hspace{-1.cm}
    \vspace{-0.3cm}
    \caption{    \label{fig:1d_sde_inverse_u}
    sPI-GeM for  inverse SDE  problem: (a) the eigenvalues of the covariance matrix for predicted $u$, predicted (b) mean, and (c) standard deviation for $u$. Blue dashed lines in (b): Locations of the sensors for $u$. }
\end{figure}

\begin{figure}[H]
    \centering
    \subfigure[]{
    \includegraphics[width=0.36\textwidth]{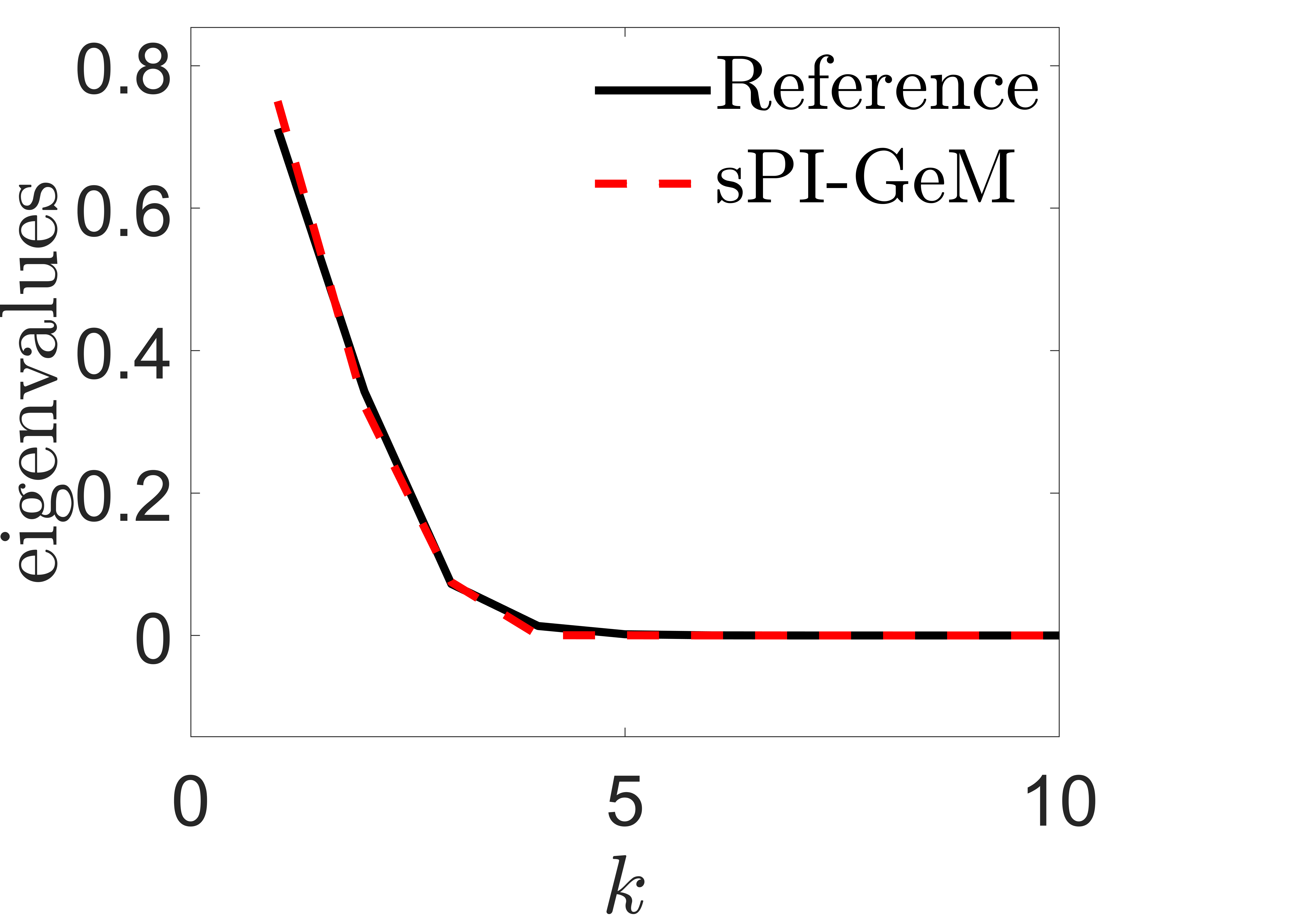}}\hspace{-1.cm}
    \subfigure[]{
    \includegraphics[width=0.36\textwidth]{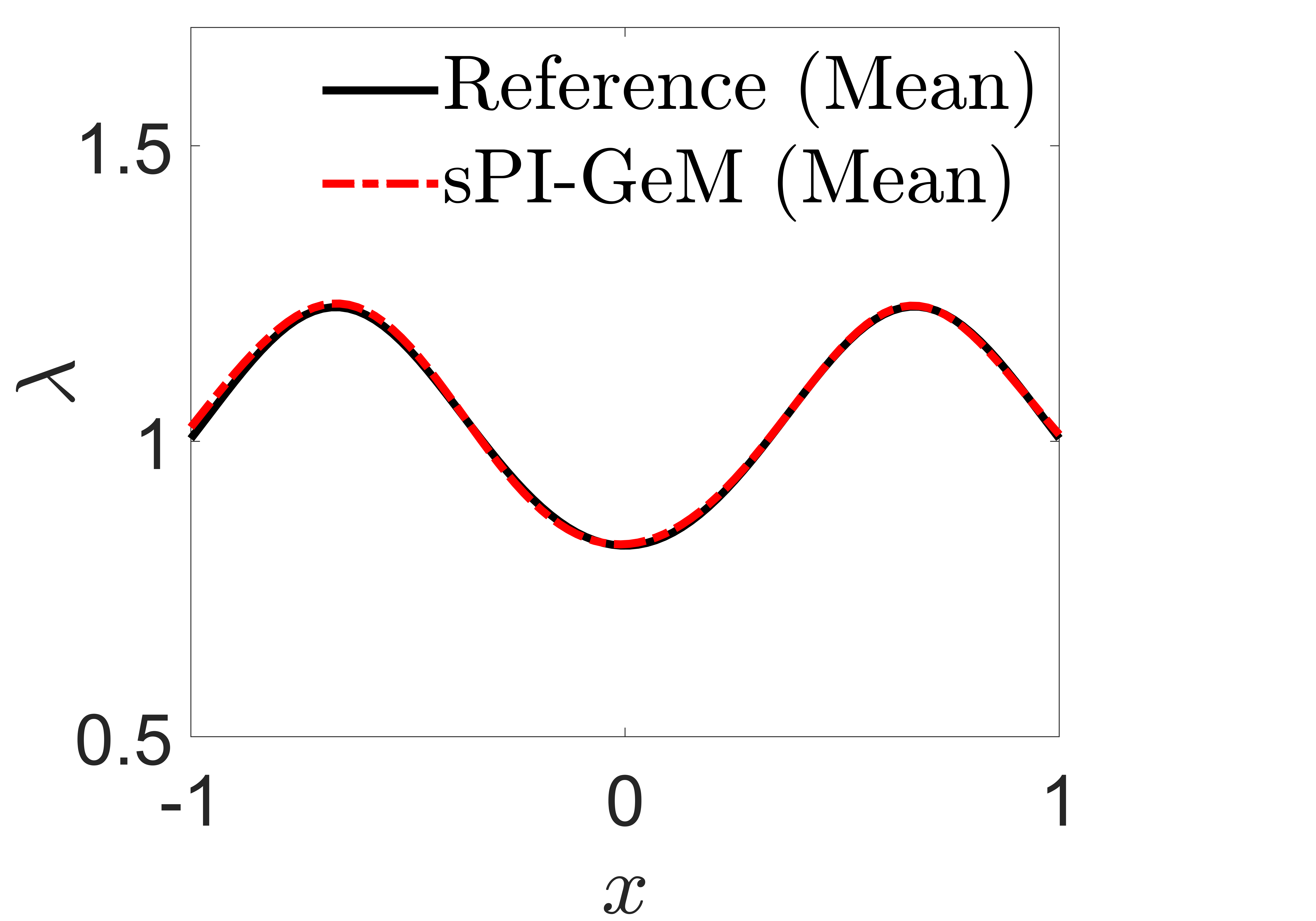}}\hspace{-1.cm}
    \subfigure[]{
    \includegraphics[width=0.36\textwidth]{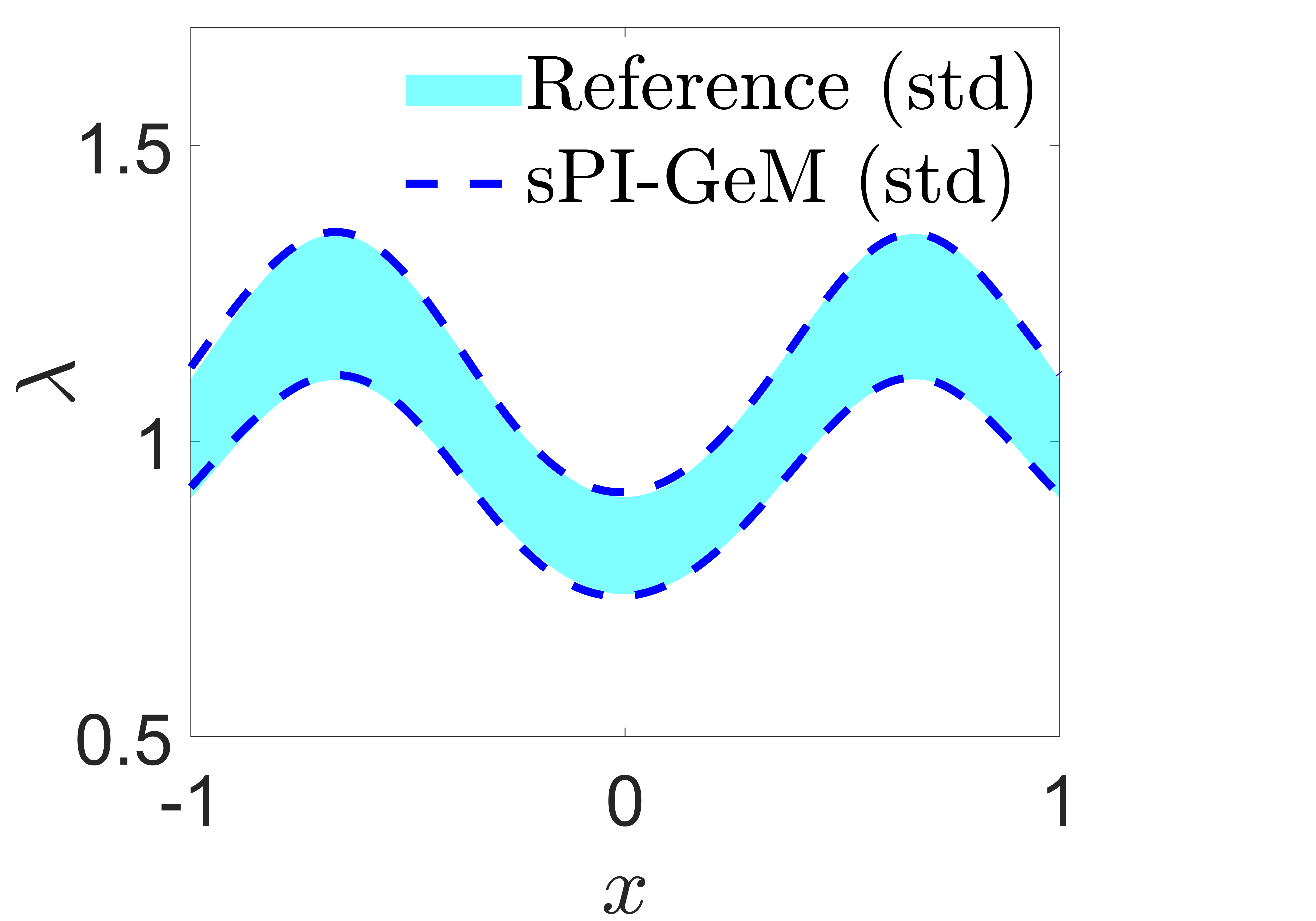}}\hspace{-1.cm}
    \vspace{-0.3cm}
    \caption{\label{fig:1d_sde_inverse_lambda}
    sPI-GeM for  inverse SDE  problem: (a) the eigenvalues of the covariance matrix for predicted $\lambda$, predicted (b) mean, and (c) standard deviation for $\lambda$. All the results of sPI-GeM are computed based on 10,000 generated samples for $\lambda$.}
\end{figure}

\begin{figure}[H]
    \centering
    \includegraphics[width=0.6\textwidth]{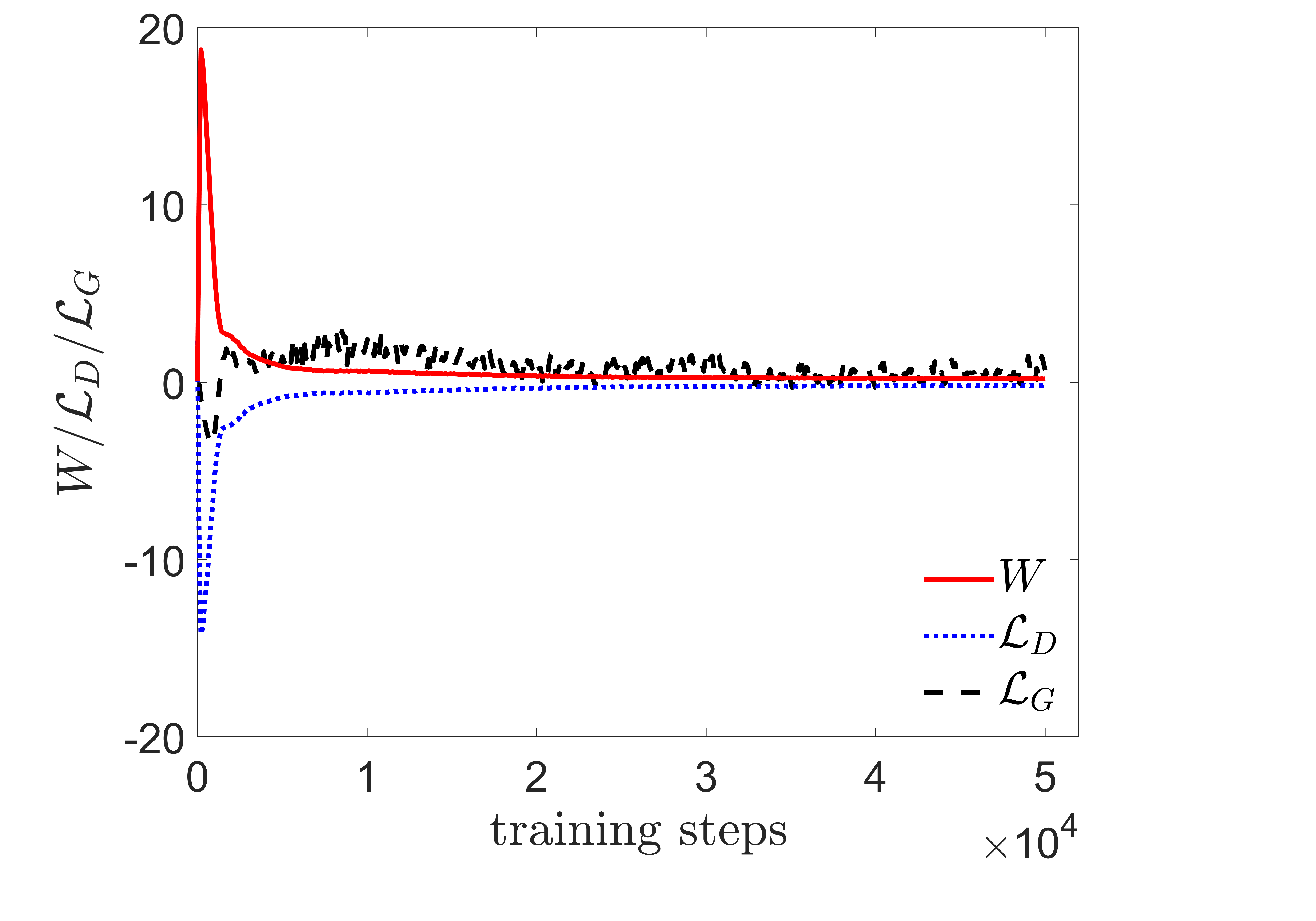}
    \caption{\label{fig:1d_sde_inverse_loss}
    sPI-GeM for inverse SDE  problem: Loss history for sPI-GeM. $\mathcal{L}_D$: Loss for the discriminator; $\mathcal{L}_G$: Loss for the generator; $\bm{W}$: Estimation of the Wasserstein-1 distance.}
\end{figure}

With the  trained PI-BasisNet, we can obtain the coefficients for $u$ and $\lambda$ used in the training data i.e. $\bm{\psi}_u$ and $\bm{\psi}_{\lambda}$. We then train GeM to approximate the distributions for $\bm{\psi}_u$ and $\bm{\psi}_{\lambda}$. Further, we can generate samples for (1) $u$ and $\lambda$, and (2) $f$ based on the generated $u$ and $\lambda$ as well as  Eq. \eqref{eq:output_sgem}. Specifically,
we generate 10,000 samples for $u$, $\lambda$ and $f$ from the trained sPI-GeM. Each sample is resolved by 100 uniform discrete points.  The eigenvalues of the covariance matrix, and the predicted mean/std for $u$ and $\lambda$ are illustrated in Figs. \ref{fig:1d_sde_inverse_u} and \ref{fig:1d_sde_inverse_lambda}, respectively. It is observed that all the results are in good agreement with the reference solutions. We note that the reference solution for (1) $u$ is computed using 10,000 samples generated from the finite difference method \cite{psaros2023uncertainty}, and (2) $\lambda$ is obtained using 10,000 samples generated from Eq. \eqref{eq:inverse_sde_f}.  We also present the computational errors for $u$, $\lambda$ and $f$ in Table \ref{tab:1d_sde_inverse_tab}, which again demonstrates the good accuracy of the sPI-GeM for solving inverse SDE problem. Finally, the loss history of the sPI-GeM in Fig. \ref{fig:1d_sde_inverse_loss} indicates that the present approach converges in around 20,000 training steps for this particular case.

\begin{table}[H]
\centering
\caption{sPI-GeM for inverse SDE problem: relative errors for $u$,~$\lambda$, and $f$.}
\begin{tabular}{lcccc}
\toprule
& \textbf{\thead{predict mean\\Rel. $L_2$ Error}}  & \textbf{\thead{predict std \\Rel. $L_2$ Error}}\\
\midrule
$u$ &  $1.11  \%$ & $0.70 \%$\\
$\lambda$ &  $0.56 \%$ & $2.40\%$\\
$f$ &  $1.86 \%$ & $1.64\%$\\
% $\bm{f}$ & $9.60\times10^{-3}$ & $4.91\times10^{-4}$ & $1.89\times10^{-2}$ & $9.66\times10^{-4}$\\
\bottomrule
\label{tab:1d_sde_inverse_tab}
\end{tabular}
\end{table}
% \color{red}

\vspace{-1.cm}

\subsection{High-dimensional SDE problem ($D_{\bm{x}} = 20$)}
\label{sec:sine_gordon}

We proceed to solve a nonlinear stochastic Sine–Gordon equation with 20 dimensions in the spatial domain, which is expressed as:  
\begin{equation}
    \Delta u(\bm{x},\bm{\zeta}) + \text{sin}(u(\bm{x}, \bm{\zeta})) = f(\bm{x}, \bm{\zeta}), \quad \bm{x} \in \mathbb{B}^{D_{\bm{x}}},
    \label{eq:Sine–Gordon}
\end{equation}
where $\Delta$ is the Laplacian operator, $u(\bm{x},\bm{\zeta})$ represents the solution, $f(\bm{x},\bm{\zeta})$ is the source term, and $\mathbb{B}^{D_{\bm{x}}}$ denotes a $D_{\bm{x}}$-dimensional unit sphere where $D_{\bm{x}} = 20$. Here we would like to solve Eq. \eqref{eq:Sine–Gordon} given snapshots on $f(\bm{x}, \bm{\zeta})$ and the boundary conditions using sPI-GeM. Specifically, the exact solution to Eq. \eqref{eq:Sine–Gordon} in this study is expressed as \cite{hu2024tackling}:
% \begin{equation}
%     u_{\text{exact}}(\bm{x}, \bm{\zeta}) = \left(1 - \|\bm{x}\|_2^2\right) \left(\sum_{i=1}^{d-1} \zeta_i \sin(x_i + \zeta_{i+d-1}\cos(x_{i+1}) + x_{i+1} \cos(x_i)) + \mathcal{F}(x_0)\right),
%     \label{eq:u_sine_gordon}
% \end{equation}
\begin{align}
    u_{\text{exact}}(\bm{x}, \bm{\zeta}) 
    &= \left(1 - \|\bm{x}\|_2^2\right) \Bigg(
        \sum_{i=1}^{d-1} \zeta_i \sin\big(
            x_i + \zeta_{i+d-1}\cos(x_{i+1}) \notag \\
    &\quad + x_{i+1} \cos(x_i)
        \big) + \mathcal{F}(x_0)
    \Bigg)
    \label{eq:u_sine_gordon}
\end{align}
where $d = 20$, $\zeta_i/\zeta_{i+d-1} \sim N(0.5, 1)$, and  $\mathcal{F}(x_0)$ is a truncated Karhunen–Loève expansion  for a specific Gaussian process \cite{solin2020hilbert}:
\begin{subequations}\label{eq:inverse_sde_f}
\begin{align}
    &\mathcal{GP}(x_0) \approx \mathcal{F}(x_0) =  \sum_{n=0}^{N} \sqrt{\lambda_n} \, \xi_n \, \phi_n(x_0), \\
    &\phi_0(x_0) = 1, \quad \phi_n(x_0) =  \cos(n \pi \frac{x_0+1}{2}),\\
    &\lambda_n = \frac{\sigma^2 \sqrt{2\pi} l}{2} \exp\left( -\frac{1}{2} \left( \frac{n\pi l}{2} \right)^2 \right),
\end{align}
\end{subequations}
where $N = 50$, $l = 0.08$, $\sigma = 1$, and $\xi_n \sim N(0,1)$ are independent standard Gaussian random variables. Here, $\lambda_n$ and $\phi_n(x)$ denote the eigenvalues and the corresponding orthonormal basis functions, respectively, obtained from the spectral decomposition of the squared exponential kernel with correlation lengths $l = 0.08$. The boundary conditions and $f(\bm{x}, \bm{\zeta})$ can then be derived based on Eqs. \eqref{eq:Sine–Gordon} and \eqref{eq:u_sine_gordon}. We note that in this particular case, (1) we have more than 50 dimensions in the stochastic domain, and (2) the solution $u$ exhibits more pronounced fluctuations along the $x_0$ direction, demonstrating heterogeneous behaviors in different dimensions of the spatial domain.

For the test case considered here, we assume that we have 5,000 snapshots for $f$ and we have sufficient measurements for each snapshot. The boundary conditions are hard encoded in $\mathcal{NN}_B$ as in \cite{hu2024tackling}. Similarly, we first train the PI-BasisNet to obtain the coefficients as well as the basis functions for $u$. To reduce the computational cost, we use the dimension reduction technique discussed in Sec. \ref{sec:method} to obtain the input for $\mathcal{NN}_C$ in PI-BasisNet. Specifically, we randomly select 1024 locations in the spatial domain, and assume that we have sensors at these locations for each snapshot of $f$. We then perform the principal component analysis (PCA) to these selected data of $f$.  The coefficients of the first 64 modes are employed as the representation for $f$ and thus the input for $\mathcal{NN}_C$. The output of BasisNet is the approximation to $u$. As we encode Eq. \eqref{eq:Sine–Gordon} in PI-BasisNet, we can then obtain the predictions for $f$. In the training of PI-BasisNet, we randomly select 128 points for $f$ at each training step to compute the residue of the equation for computational efficiency.

Similar as in the previous cases, we can obtain $\bm{\psi}_u$ from the trained PI-BasisNet, which is used to train GeM to learn the distribution over the coefficients of $u$. Samples on $u$ and $f$, can be obtained following Eq. \eqref{eq:output_sgem} when the PI-GeM is trained. 
To demonstrate the accuracy of the present model for solving SDE problem in high-dimensional spatial space, we randomly select 10,000 locations in the 20-dimensional spatial space, and predict $u/f$ using the sPI-GeM at these locations. The computational errors for the predictions are illustrated in Table \ref{tab:Sine–Gordon_mean_tab}.  We can see that the relative $L_2$ errors for the predicted mean and standard deviations of $u/f$ are smaller than $5\%$, demonstrating that the present approach is capable of achieving good accuracy for solving SDEs with both high-dimensional stochastic and spatial space.

\begin{table}[ht]
\centering
\caption{sPI-GeM for high-dimensional SDE problem ($D_{\bm{x}} = 20$):  relative $L_2$ errors for $u$ and $f$.}
\begin{tabular}{lcccc}
\toprule
& \textbf{\thead{predict mean\\Rel. $L_2$ Error}}  & \textbf{\thead{predict std \\Rel. $L_2$ Error}}\\
\midrule

$u$ & $3.04\%$  & $2.88\%$ \\
$f$ & $3.17\%$  & $2.93\%$ \\

\bottomrule
\label{tab:Sine–Gordon_mean_tab}
\end{tabular}
\end{table}

\begin{figure}[H]
    \centering
    \includegraphics[width=0.6\textwidth]{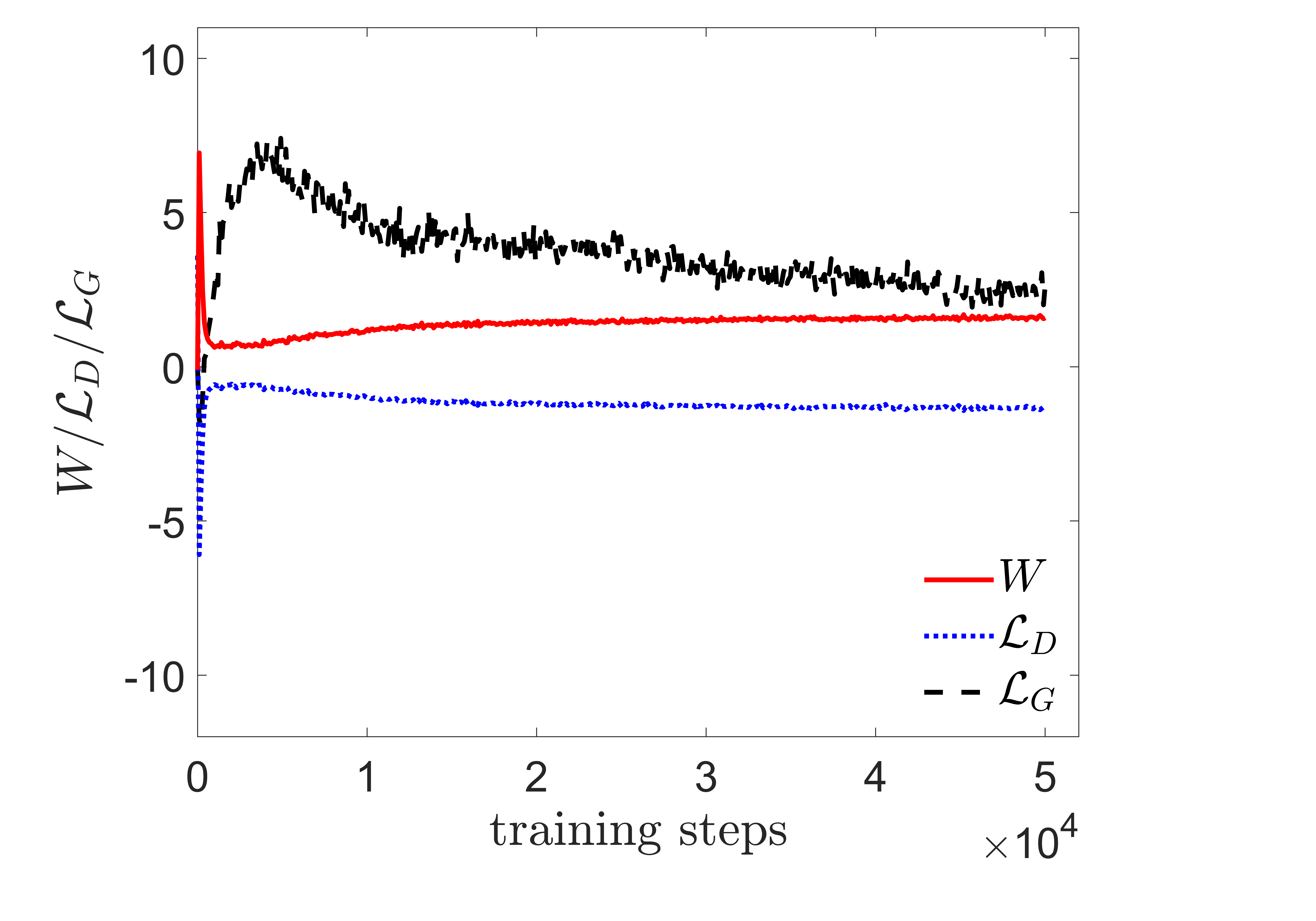}
    \caption{sPI-GeM for high-dimensional SDE problem ($D_{\bm{x}} = 20$): Loss history for sPI-GeM. $\mathcal{L}_D$: Loss for the discriminator; $\mathcal{L}_G$: Loss for the generator; $\bm{W}$: Estimation of the Wasserstein-1 distance.}
    \label{fig:Sine–Gordon_mean_loss}
\end{figure}

We finally present the loss history for sPI-GeM in Fig. \ref{fig:Sine–Gordon_mean_loss}. As shown, the discriminator in sPI-GeM converges in around 20,000 training steps, which is quite efficient since the SDE problem consider here is highly nonlinear and has both high-dimensional stochastic and spatial space.

\section{Summary}
\label{sec:summary}
We developed a novel scalable physics-informed deep generative model (sPI-GeM) for solving forward and inverse stochastic differential equations given data. Unlike the existing deep learning models for solving the SDE problems that are only scalable in the stochastic space, the sPI-GeM is capable of solving SDE problems with both high-dimensional stochastic and spatial space. The sPI-GeM is composed of two deep learning models, i.e., (1) physics-informed basis networks (PI-BasisNet), which are to learn the basis functions as well as the coefficients given data on a certain stochastic process, and (2) physics-informed deep generative model, which learns the distribution over the coefficients obtained from the PI-BasisNet. New samples for the learned stochastic process can then be obtained via the inner product between the output of the generator and the basis functions from the trained PI-BasisNet. We first employed the proposed approach to approximate Gaussian and non-Gaussian stochastic processes. The results showed that the sGeM is able to achieve good accuracy and also converges faster as compared to the existing deep generative models for learning the same stochastic process.  
We also performed numerical experiments on solving forward and inverse SDE problems using the sPI-GeM to show the good accuracy of the proposed method.  In particular,  we demonstrated the scalability of the sPI-GeM in both the stochastic and spatial space using an example of a forward SDE problem with both high-dimensional stochastic ($> 50$ dimensions) and spatial (20 dimensions) space, which has not been reported in the existing work to the best of our knowledge.

In the current study, we only focus on the steady SDE problems, but we note that sPI-GeM can be readily extended to unsteady SDE problems. Also, the sPI-GeM is quite flexible and can be seamlessly integrated with other models. For instance, (1) we can replace the GANs with other deep generative models that are easier to train, e.g., diffusion model \cite{ho2020denoising,croitoru2023diffusion}, flow matching \cite{lipman2022flow}, and score-based generative model \cite{song2020score}; and (2) we can employ the stochastic dimension gradient descent developed in \cite{hu2024tackling} for the training of PI-BasisNet if we would like to tackle SDE problems with even higher dimensional spatial space.  {Also,  we acknowledge that most of the test cases in the present work are synthetic. Applications of the proposed method to more realistic physical scenarios, such as solving the random Schr$\ddot{\mbox{o}}$dinger equation in disordered solids, remains an important direction for future research. Furthermore, although the sGeM/PI-sGeM achieved good accuracy in all the numerical experiments, rigorous analyses on the convergence as well as generalization errors of the present method for solving forward and inverse SDEs are of interest in future work.}

\section*{Acknowledgements}
S. Z., W. Y., and X. M. acknowledge the support of the National Natural Science Foundation of China (No. 12201229) and the Interdisciplinary Research Program of HUST (No. 2024JCYJ003). X. M. also acknowledges the support of the Xiaomi Young Talents Program.  L. G. acknowledges the support of the National Natural Science Foundation of China (No. 92270115, 12071301). X. M. thanks Dr. Zongren Zou from California Institute of Technology for the helpful discussion on learning basis functions using deep neural networks.

\appendix

\section{Details on the computations}
\label{sec:computations}

In all the cases discussed in Sec. \ref{sec:results}, the Adam optimizer is utilized for training all neural networks.  For training PI-BasisNets in Sec. \ref{sec:results}, $\beta_1 = 0.9$, and $\beta_2 = 0.9$ are used in the Adam optimizer. For training the PI-GeMs, $\beta_1 = 0.5$, and $\beta_2 = 0.9$. Additional details regarding the architectures and training steps for PI-BasisNet and PI-GeMs are provided in Tables  \ref{table:deeponet_arch} - \ref{table:batch_size_points}. The CNN in Sec. \ref{sec:high_d_zeta} has three convolutional layers, each followed by ReLU activation function and a \(2 \times 2\) max-pooling layer. The number of features in each layer is 16, 32, and 64, respectively. The output of the last convolutional layer is then flattened and fed to a  fully-connected neural network (FNN) with three hidden layers and 64 neurons each. The activation function in the FNN is also ReLU. In addition, the tensor neural networks (TNNs) \cite{wang2022tensor} are used for $\mathcal{NN}_B$ in the PI-BasisNet for computational efficiency. In each TNN, the input dimension is one and the output dimension is 64. We therefore have two TNNs  for $\mathcal{NN}_B$ since the number of spatial dimensions is two here. Also, for the case with feature expansion we use  $[\bm{x},\text{sin}(\bm{x}),\text{cos}(\bm{x}),\cdots, \text{sin}(13\bm{x}),\text{cos}(13\bm{x})]$ as the input for $\mathcal{NN}_B$ instead of  $\bm{x}$.
In Sec. \ref{sec:inverse_pde}, we employ two neural networks to learn the basis functions for $u$ and $\lambda$, respectively.  Further, we utilize only one neural network for $\mathcal{NN}_C$ with the output dimension equal to 128. We divide the outputs into two parts. The first half  is for approximating the coefficients of $u$, and the remainder are the coefficients for $\lambda$. The reference solutions for Sec. \ref{sec:darcy_2d} are obtained using the {\sl MATLAB PDE Toolbox} with adaptively generated triangular meshes, in which the maximum edge length is set as $H_{\max} = 0.06$.

\begin{table}[H]
\caption{
Architecture and training steps of PI-BasisNet in each case. \(\text{lr}\) represents the learning rate.
}
\centering
\footnotesize
\renewcommand{\arraystretch}{1.2}
\setlength{\tabcolsep}{3pt} 
\begin{tabular}{l l l l l l c}
\toprule
  & \multicolumn{2}{l}{$\mathcal{NN}_C$}  & \multicolumn{2}{l}{$\mathcal{NN}_B$} & \multirow{2}{*}{\thead{Training\\steps}}& \multirow{2}{*}{lr}  \\
  \cmidrule(lr){2-3} \cmidrule(lr){4-5}
  & width $\times$ depth  & Activation & width $\times$ depth & Activation & &\\
\midrule
  {Sec. \ref{sec:part_1}} & $128 \times 3$   & LeakyReLU & $128 \times 3$ & tanh &  50,000 &$1 \times 10^{-3}$\\
  {Sec. \ref{sec:high_d_zeta}} & CNN   & ReLU & $128 \times 3$  & tanh &  50,000 &$1\times 10^{-4}$\\
  {Sec. \ref{sec:darcy_2d}} & \thead[l]{CNN($u$)\\CNN($\lambda$)}   & \thead[l]{ReLU\\ReLU} & \thead[l]{$128 \times 3$ ($u$)\\$128 \times 3$ ($\lambda$)} & \thead[l]{tanh\\tanh} &  50,000 &$1\times 10^{-3}$ \\
  {Sec. \ref{sec:inverse_pde}} & $128 \times 3$   & LeakyReLU & \thead[l]{$128 \times 3$ ($u$)\\$128 \times 3$ ($\lambda$)} & \thead[l]{tanh\\tanh} &  50,000 &$1\times 10^{-4}$ \\
  {Sec. \ref{sec:sine_gordon}} & $128 \times 3$   & LeakyReLU & $128 \times 3$ & tanh &  20,000 &$1\times 10^{-3}$\\
\bottomrule
\end{tabular}
\label{table:deeponet_arch}
\end{table}

\begin{table}[H]
\caption{Architecture and training steps of PI-GeM in each case. The width and depth are for the hidden layers. $D_{\bm{\xi}}$ represents the dimensionality of the input \(\xi\) for the generator, while \(S_D\) denotes the number of training steps for the discriminator performed for each iteration of the generator. The Training steps column refers specifically to the total number of iterations performed for the generator ($G$), and $\text{lr}$ represents the learning rate.}
\centering
\footnotesize
\renewcommand{\arraystretch}{1.2}
\setlength{\tabcolsep}{1.5pt} 
\begin{tabular}{l l l l l l l l c}
\toprule
& \multicolumn{2}{l}{G} & \multicolumn{2}{l}{D} & \multirow{2}{*}{$D_{\bm{\xi}}$} & \multirow{2}{*}{$S_D$} & \multirow{2}{*}{\thead{Training\\steps}} & \multirow{2}{*}{lr} \\
\cmidrule(lr){2-3} \cmidrule(lr){4-5}
& width $\times$ depth & Activation & width $\times$ depth & Activation & & & & \\
\midrule
{Sec. \ref{sec:part_1}} & $128 \times 3 $ & LeakyReLU & $128 \times 3$ & LeakyReLU & 50 & 10 & 50,000 & $1 \times 10^{-4}$ \\
{Sec. \ref{sec:high_d_zeta}} & $128 \times 3 $ & LeakyReLU & $128 \times 3$ & LeakyReLU & 50 & 10 & 100,000 & $1 \times 10^{-4}$ \\
{Sec. \ref{sec:darcy_2d}} & $128 \times 3 $ & LeakyReLU & $128 \times 3$ & LeakyReLU & 50 & 10 & 100,000 & $1 \times 10^{-4}$ \\
{Sec. \ref{sec:inverse_pde}} & $128 \times 3 $ & LeakyReLU & $128 \times 3$ & LeakyReLU & 50 & 10 & 50,000 & $1 \times 10^{-4}$ \\
{Sec. \ref{sec:sine_gordon}} & $128 \times 3$ & LeakyReLU & $128 \times 3$ & LeakyReLU & 50 & 10 & 50,000 & $1 \times 10^{-4}$ \\
\bottomrule
\end{tabular}
\label{table:gan_arch}
\end{table}

\begin{table}[H]
\caption{Batch size and training points for each case.}
\centering
\footnotesize
\renewcommand{\arraystretch}{1.2}
\setlength{\tabcolsep}{5.5pt}
\begin{tabular}{l l l}
\toprule
  & \textbf{Batch size} & \textbf{training points for each snapshot} \\
\midrule
  {Sec. \ref{sec:part_1}} & 10,000 & 100 (equidistantly distributed in $[-1,1]$)\\
  {Sec. \ref{sec:high_d_zeta}} & 1,000 & 128 $\times$ 128 (uniform grid in $[-\pi,\pi]^2$)\\
  {Sec. \ref{sec:darcy_2d}} & 512 &  512(randomly sampled from $101 \times 101$ uniform grid in $[-1, 1]^2$)\\
  {Sec. \ref{sec:inverse_pde}} & 1,000 & 100 (equidistantly distributed in $[-1,1]$)\\
  {Sec. \ref{sec:sine_gordon}} & 1,000 & 100 (randomly selected in $\mathbb{B}^{D_{\bm{x}}}$) \\
\bottomrule
\end{tabular}
\label{table:batch_size_points}
\end{table}

\section{Study on the  convergence of sGeM for approximating stochastic process}
\label{sec:converge_study}

In this section, we take the example of approximating the non-Gaussian process to study the effect of the number of measurements in each snapshot and the number of snapshots on computational accuracy. 

\begin{figure}[H]
    \centering
    \subfigure[]{
    \includegraphics[width=0.36\textwidth]{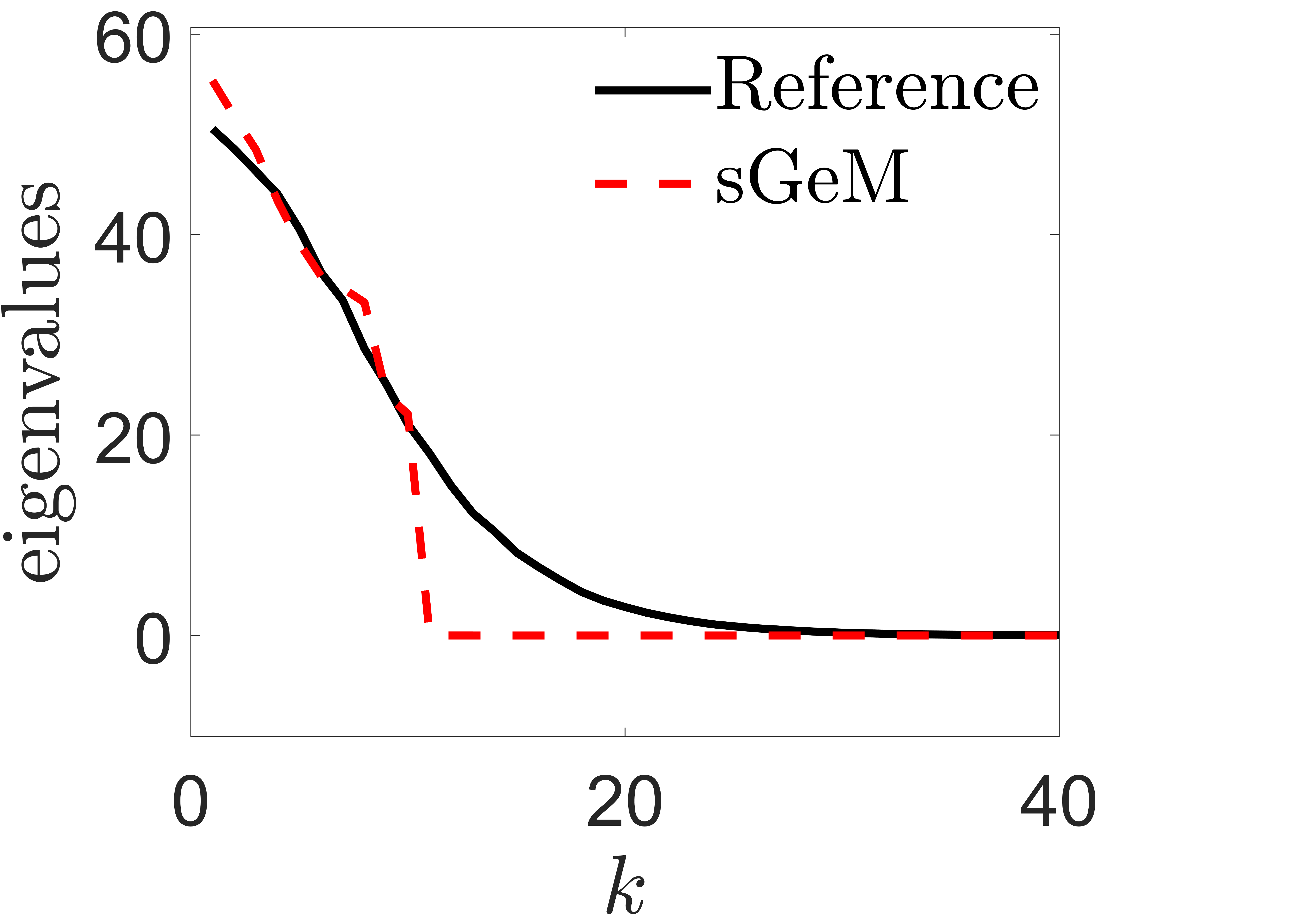}}\hspace{-1.cm}
    \subfigure[]{
    \includegraphics[width=0.36\textwidth]{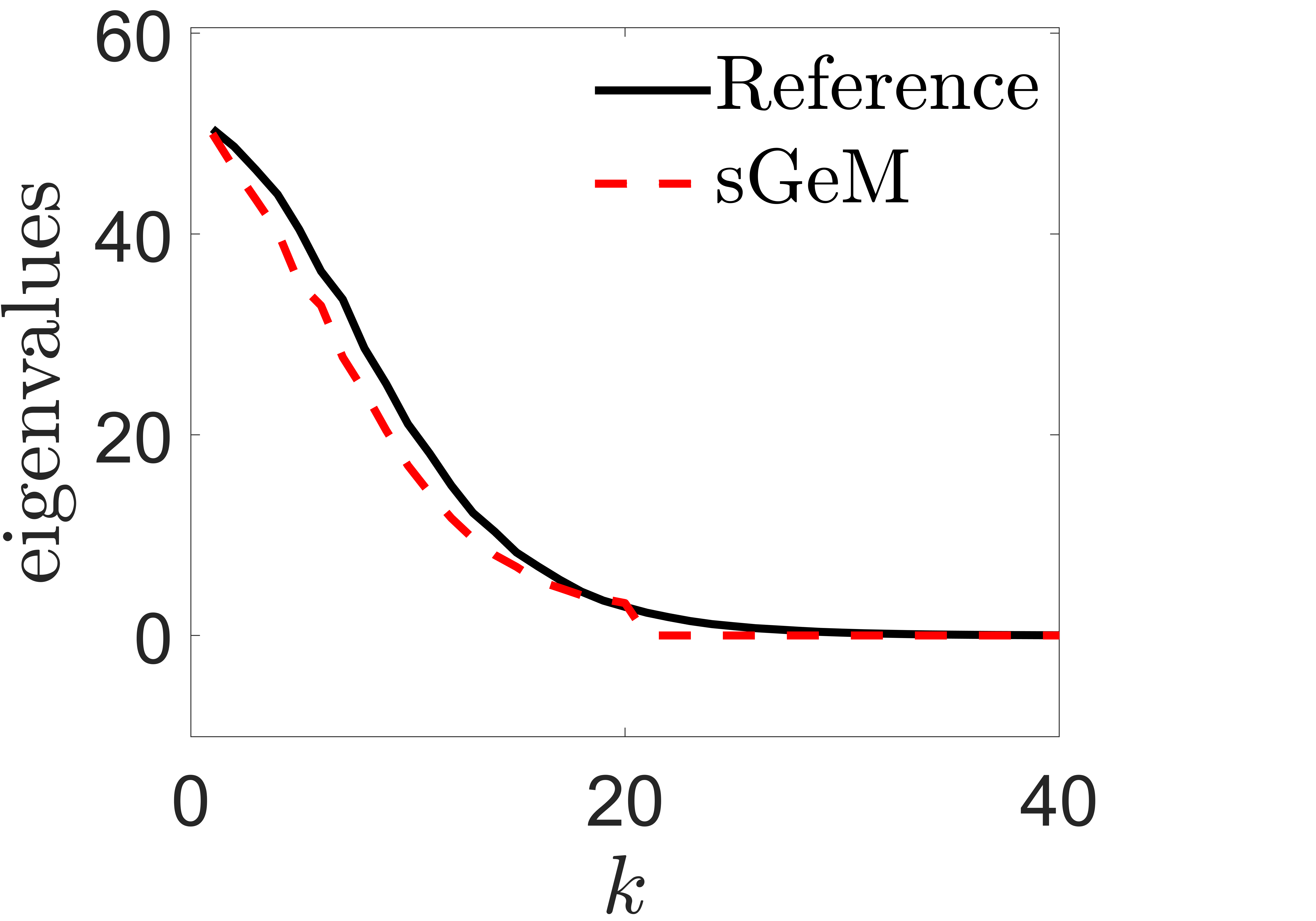}}\hspace{-1.cm}
    \subfigure[]{
    \includegraphics[width=0.36\textwidth]{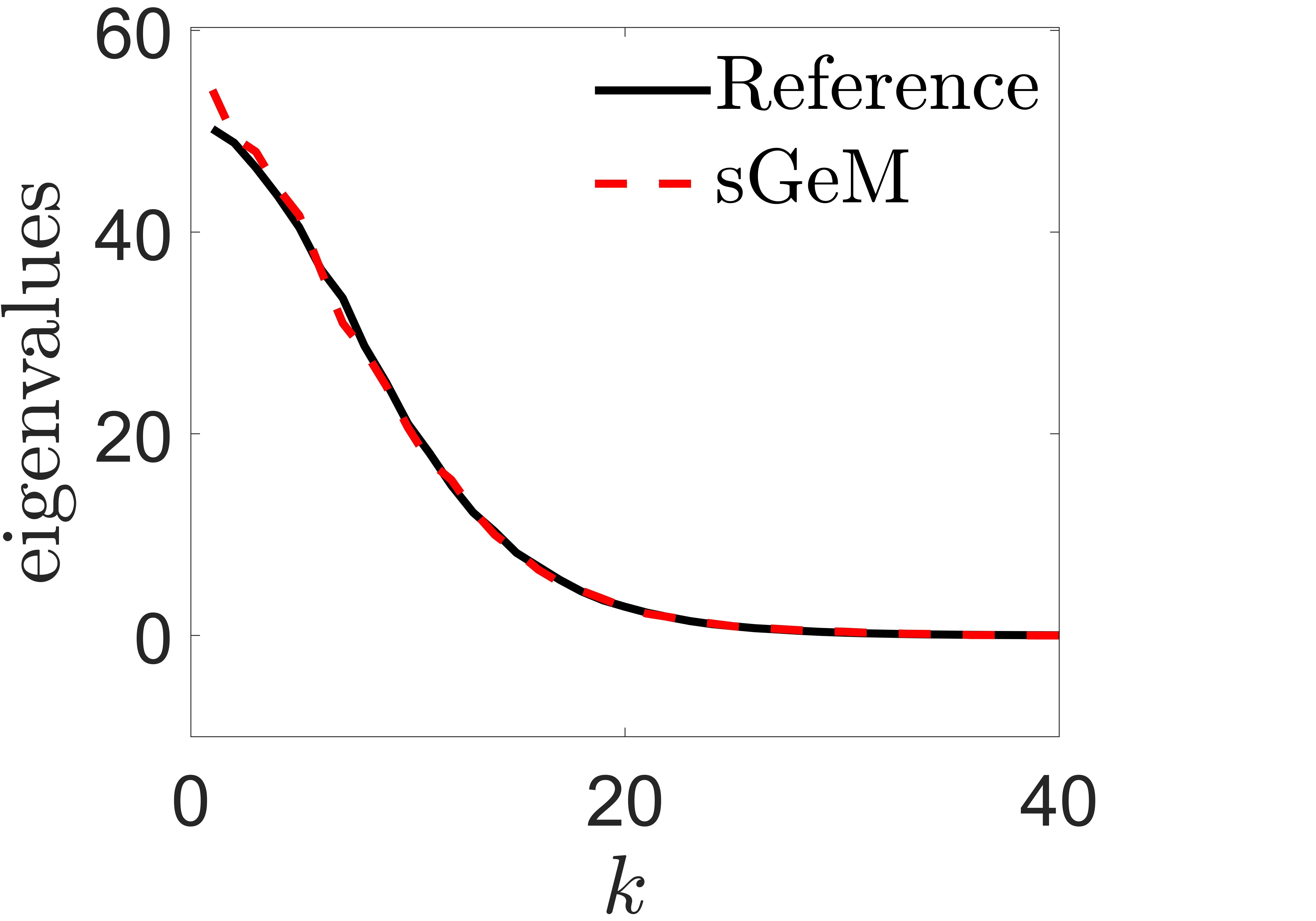}}\hspace{-1.cm}
    \vspace{-0.3cm}
    
    \subfigure[]{
    \includegraphics[width=0.36\textwidth]{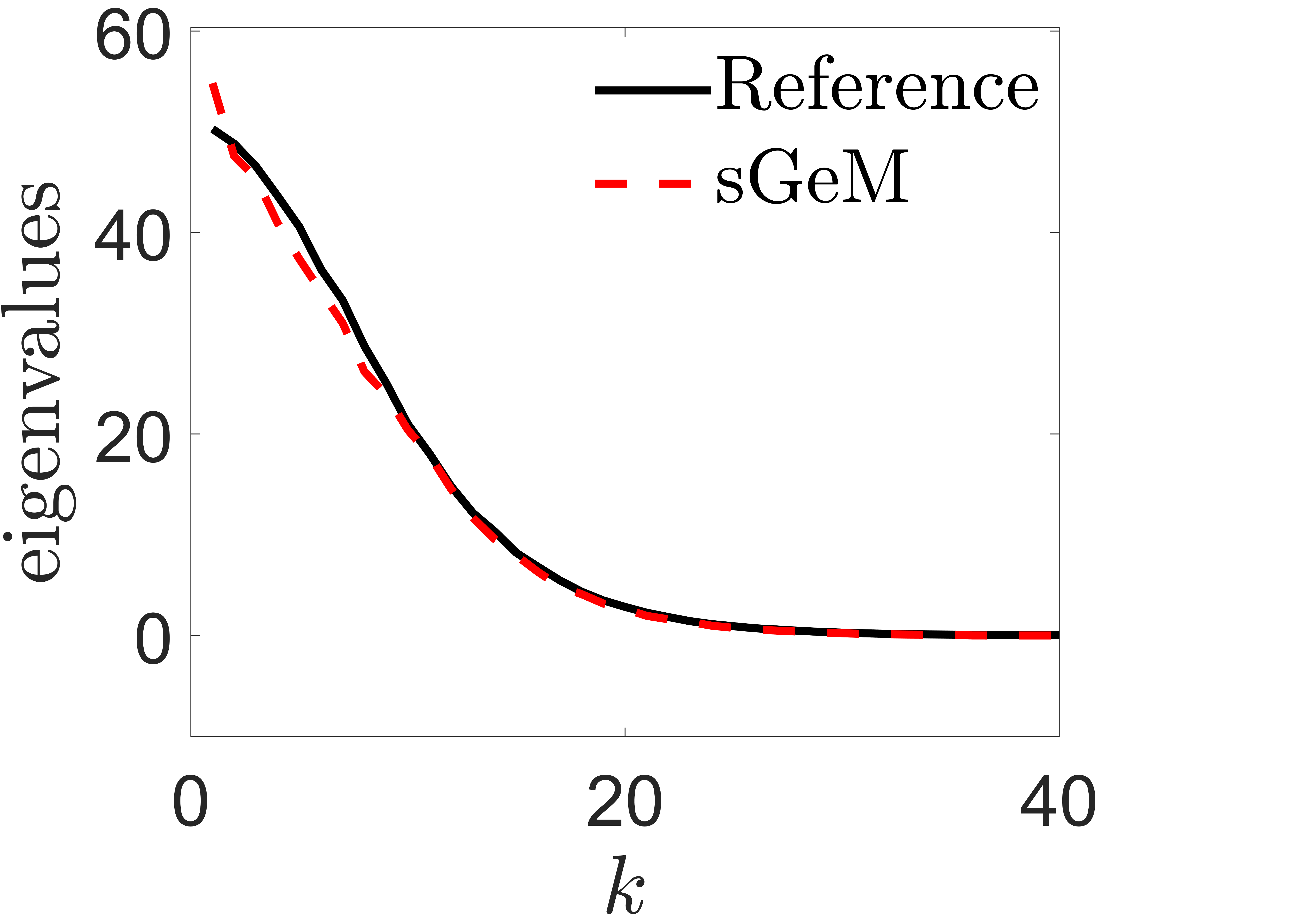}}\hspace{-1.cm}
    \subfigure[]{
    \includegraphics[width=0.36\textwidth]{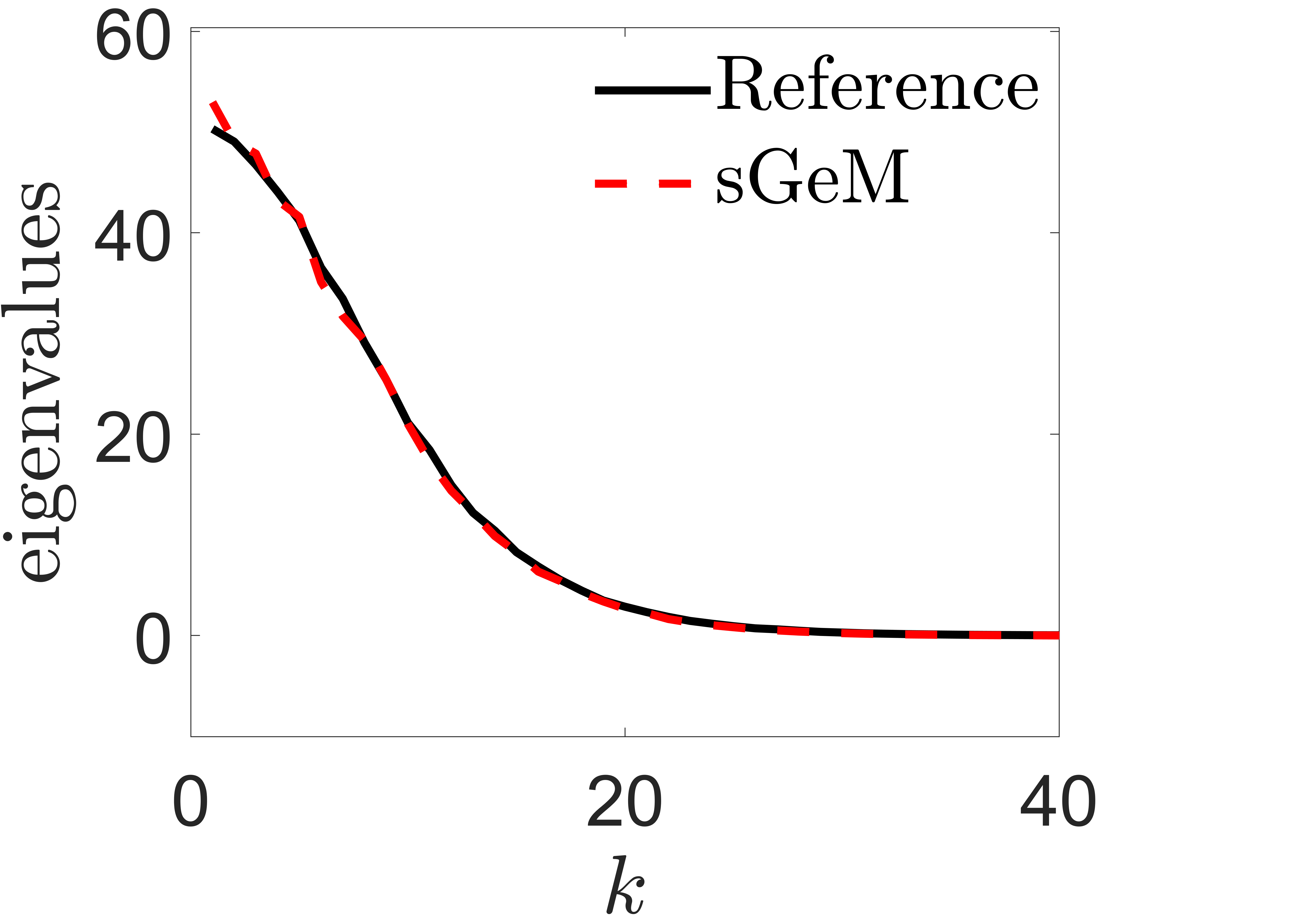}}\hspace{-1.cm}
    \subfigure[]{
    \includegraphics[width=0.36\textwidth]{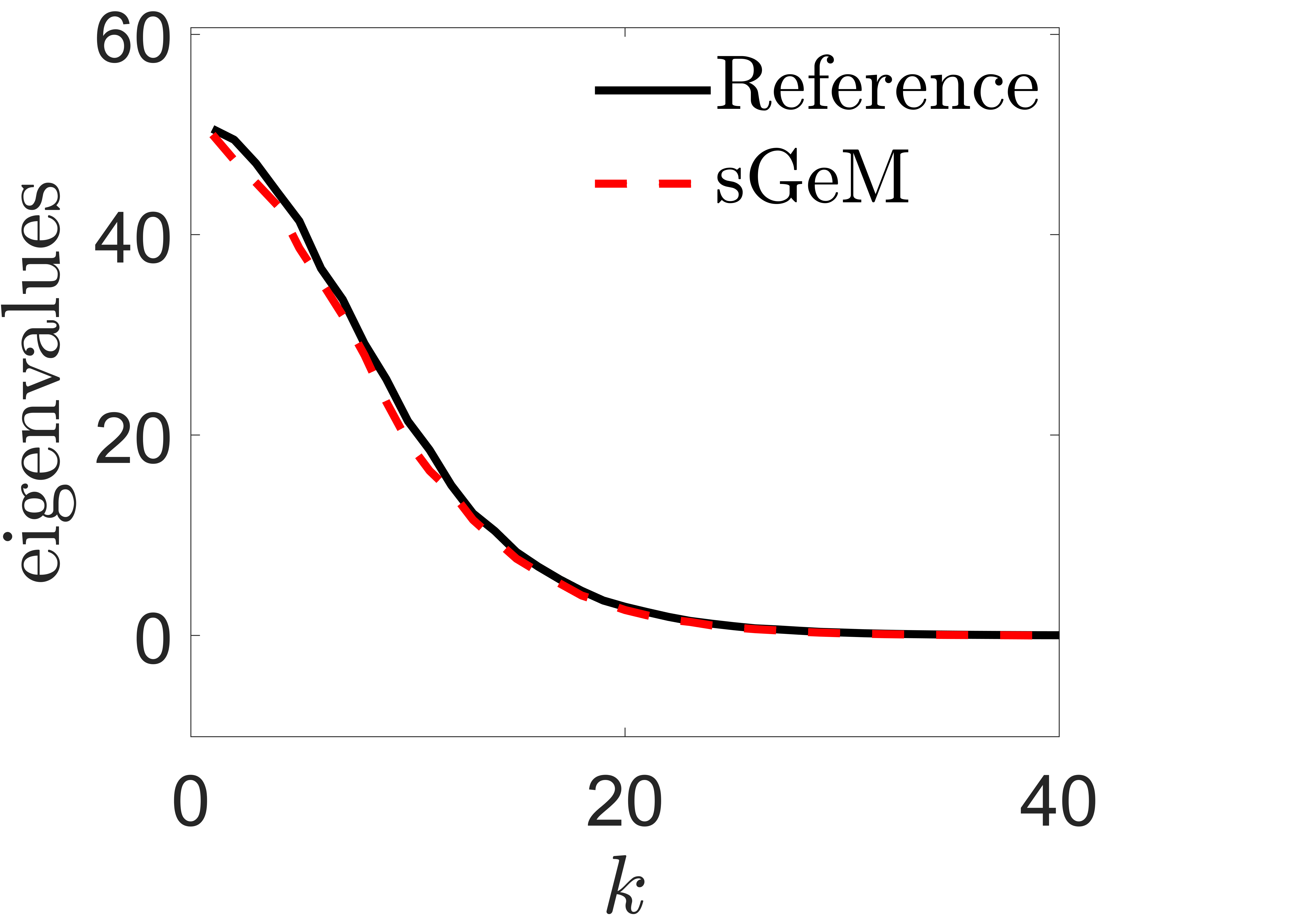}}\hspace{-1.cm}

    \caption{
    sGeM for approximating stochastic processes: Eigenvalues of the covariance matrix for predicted \(u\). (a) - (f) 10, 20, 40, 60, 80, and 100 equidistant measurements for $u$.}
\label{fig:process_test1}
\end{figure}

We first test the effect of the number of measurements in each snapshot on the computational accuracy. Specifically,  we assume that we have 10, 20, 40, 60, 80, and 100 equidistant sensors within the interval \([-1, 1]\) in each snapshot of \(u\). In addition, the number of snapshots is kept the same as in Sec. \ref{sec:part_1}. The employed architectures of the sGeM and the optimizer are also the same as in Sec. \ref{sec:part_1}. Similarly, we use the trained sGeM to generate 40,000 samples for $u$. Each sample is resolved using 100 equidistant points. The reference solution here is the same as Sec. \ref{sec:part_1}. As illustrated in Fig. \ref{fig:process_test1}, the results of sGeM show significant discrepancy from the reference solution for the case with 10 measurements in each snapshot. Good accuracy can be achieved as the number of measurements in each snapshot is larger than 20. 

\begin{figure}[H]
    \centering
    \subfigure[]{
    \includegraphics[width=0.36\textwidth]{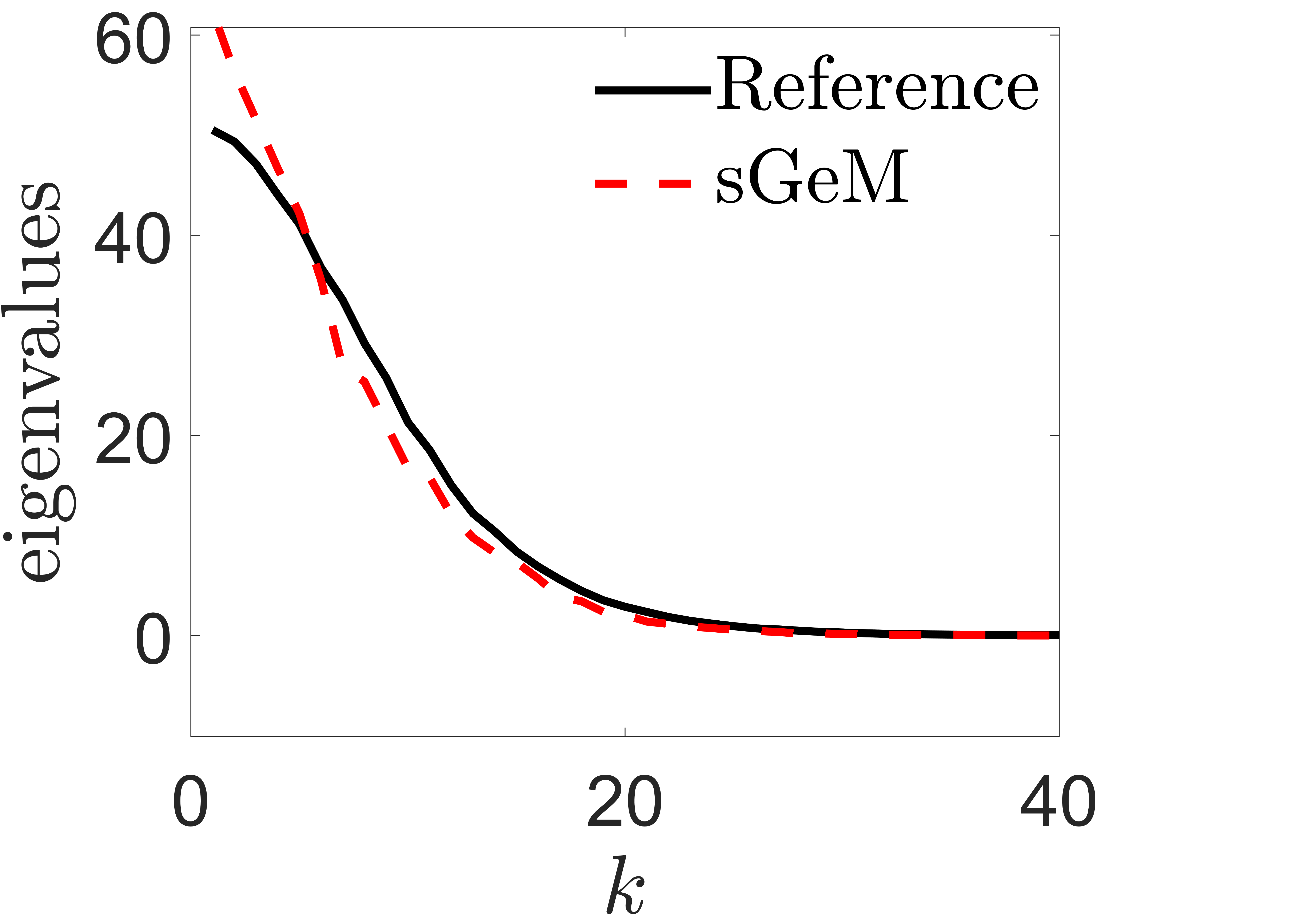}}\hspace{-1.cm}
    \subfigure[]{
    \includegraphics[width=0.36\textwidth]{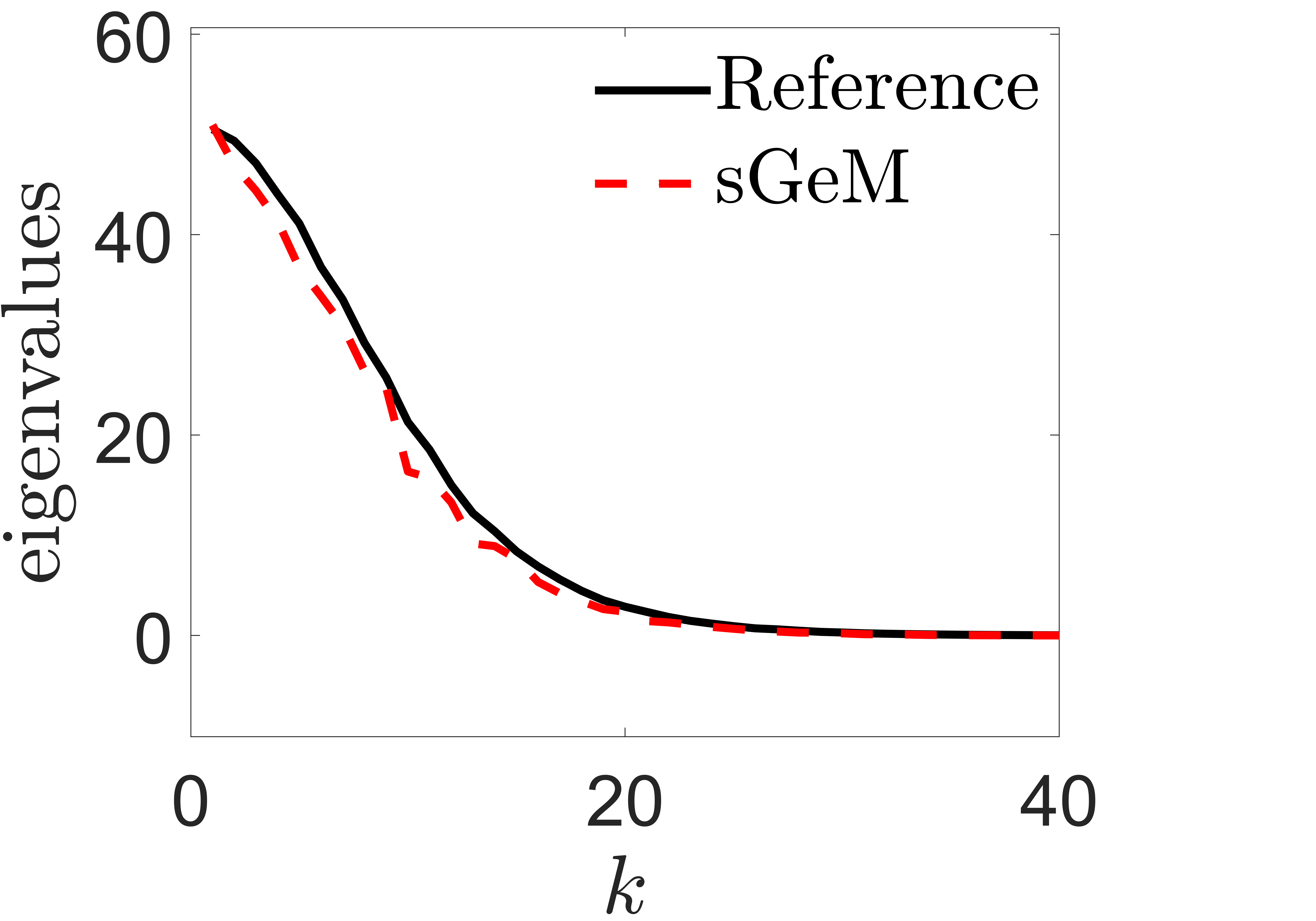}}\hspace{-1.cm}
    \subfigure[]{
    \includegraphics[width=0.36\textwidth]{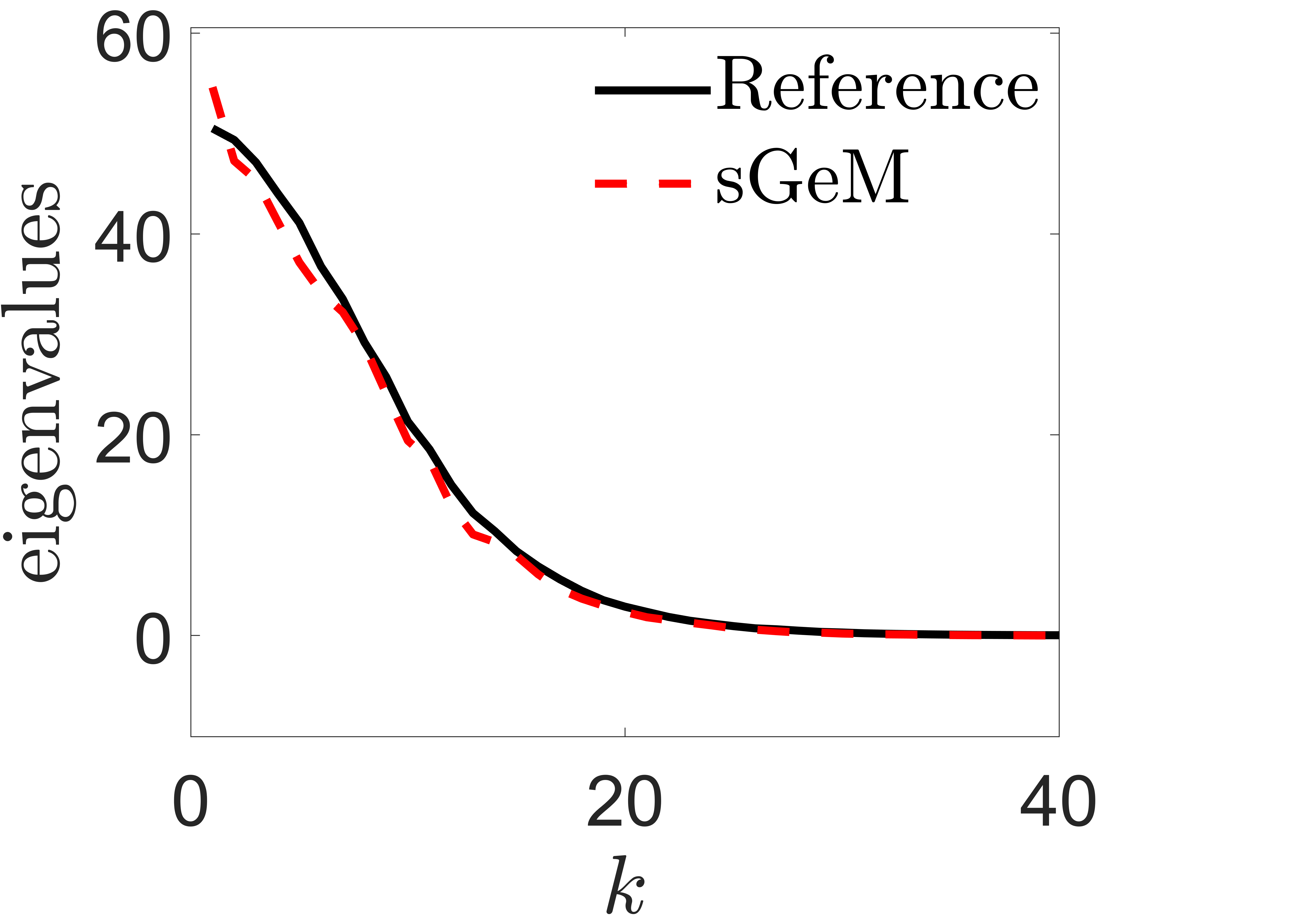}}\hspace{-1.cm}
    \vspace{-0.3cm}

    \subfigure[]{
    \includegraphics[width=0.36\textwidth]{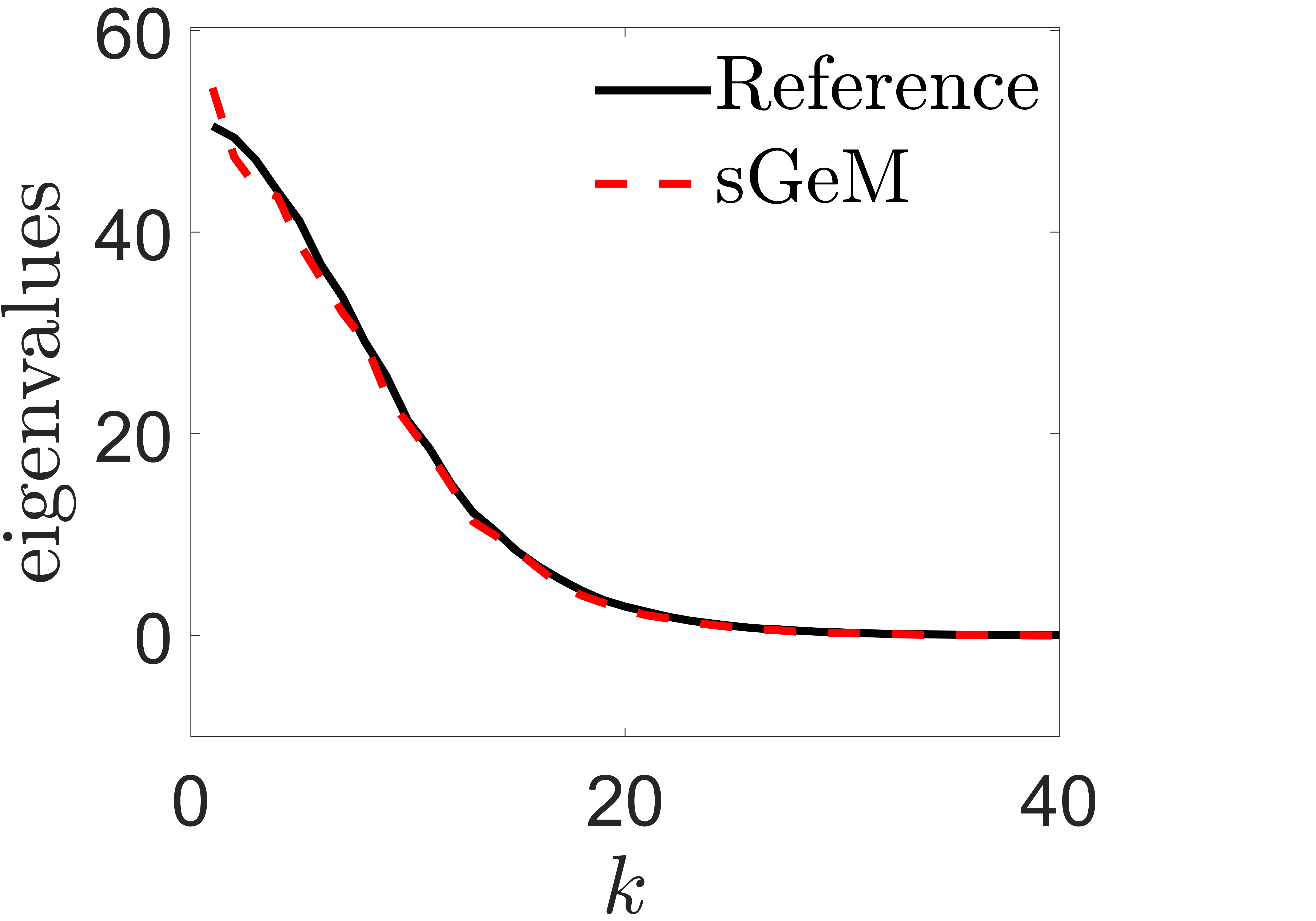}}\hspace{-1.cm}
    \subfigure[]{
    \includegraphics[width=0.36\textwidth]{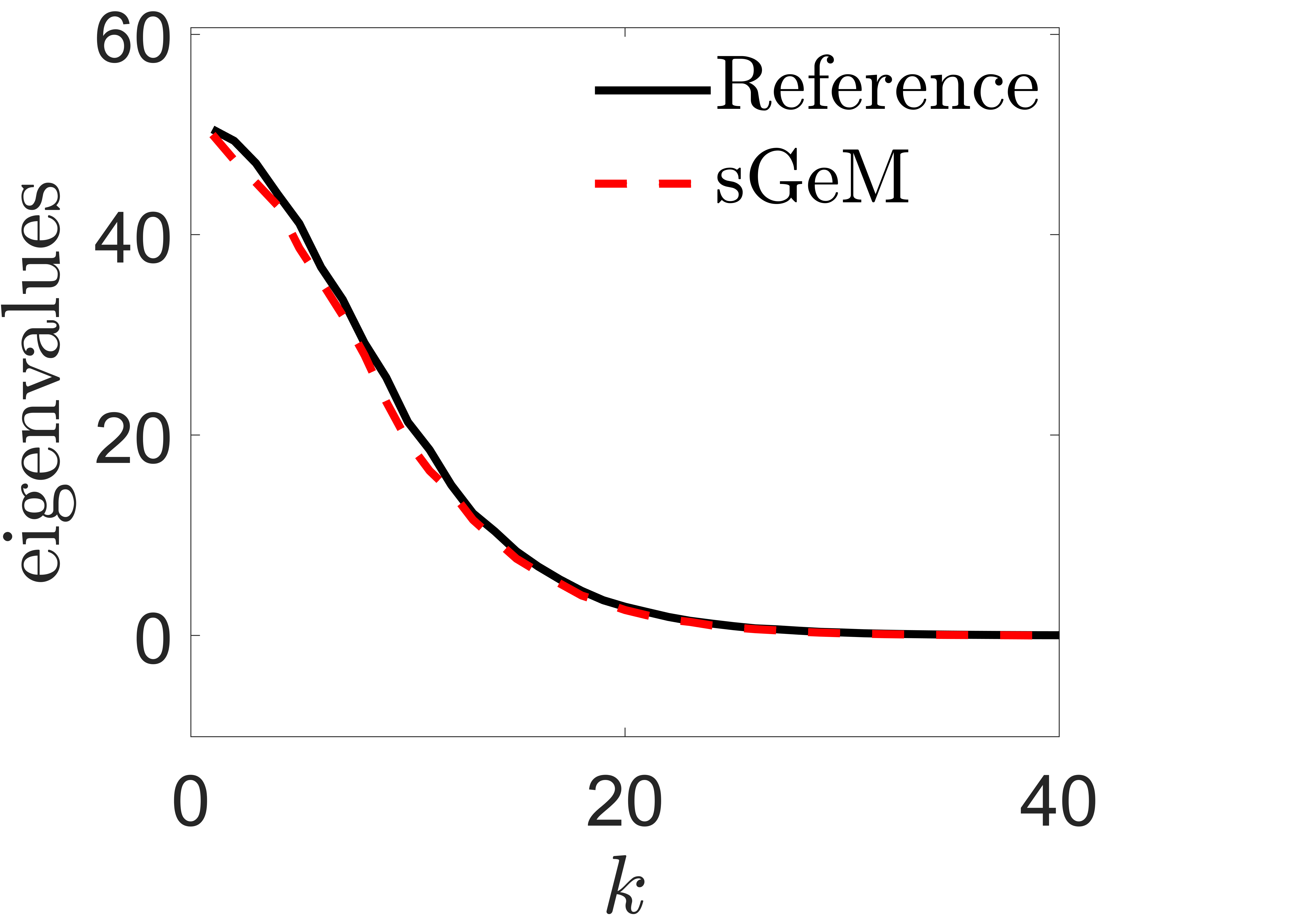}}\hspace{-1.cm}
    \subfigure[]{
    \includegraphics[width=0.36\textwidth]{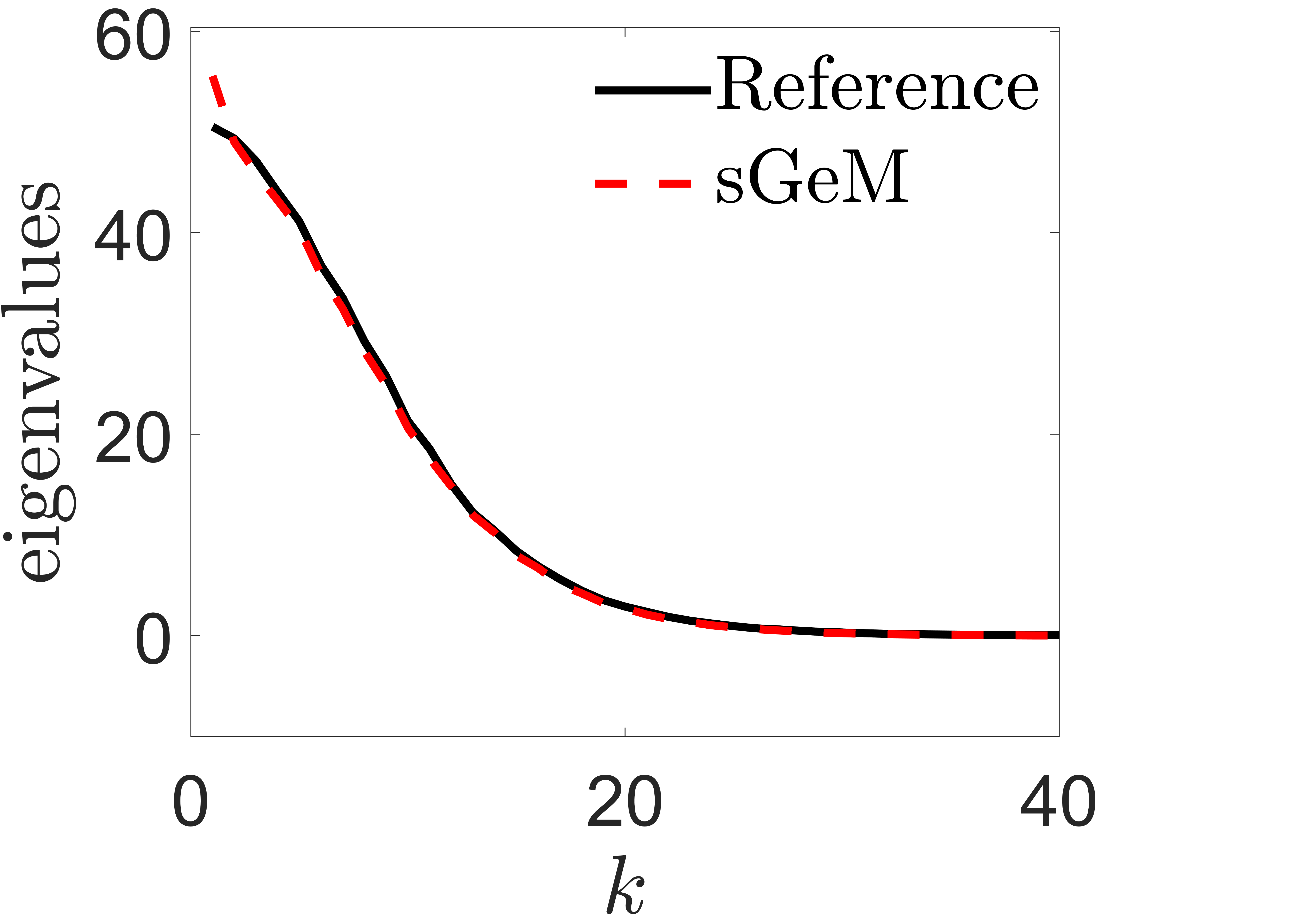}}\hspace{-1.cm}

    \caption{
    sGeM for approximating stochastic processes: Eigenvalues of the covariance matrix for predicted \(u\). (a) - (f) 500, 1,000, 2,000, 5,000, 10,000, and 40,000 samples for $u$ as the training set.}
\label{fig:process_test2}
\end{figure}

We now test the effect of the number of snapshots on the computational accuracy. Specifically,  we assume that we have 500, 1,000, 2,000, 5,000, 10,000, and 40,000 snapshots for $u$. Each snapshot is resolved by 100 equidistant sensors within the interval \([-1, 1]\) as in Sec. \ref{sec:part_1}. The employed architectures of the sGeM and the optimizer are the same as in Sec. \ref{sec:part_1}. Similarly, we also use the trained sGeM to generate 40,000 samples for $u$. Each sample is resolved using 100 equidistant points.
As we can see in Fig. \ref{fig:process_test2}, the accuracy improves as we increase the number of snapshots, and 1000 snapshots are able to provide satisfactory accuracy for this specific case.

 \bibliographystyle{unsrt} 
 \bibliography{refs}





% \end{thebibliography}
\end{document}